\documentclass[usenatbib,useAMS,manuscript]{mn2e}

\usepackage{amsmath,amsfonts,amssymb,bm,graphicx,graphics,tabularx,natbib,color}

\begin{document}

\title{\textbf{Dynamics of Elliptical Galaxies with Planetary Nebulae in Modified Newtonian Dynamics}}
\author[Tian \& Ko]
{Yong Tian$^{1}$\thanks{E-mail:yongtian@astro.ncu.edu.tw}
and Chung-Ming Ko$^{1,2}$\thanks{E-mail:cmko@astro.ncu.edu.tw}
\\
$^{1}$Institute of Astronomy, National Central University, Jhongli Dist., Taoyuan City, Taiwan (Republic of China) \\
$^{2}$Department of Physics and Center for Complex Systems, National Central University, Jhongli Dist., Taoyuan City, Taiwan (Republic of China)
}

\date{Accepted 2016 July 12. Received 2016 July 7; in original form 2016 April 30}

\maketitle

    \begin{abstract}
    The dynamics of an elliptical galaxy within a couple of effective radii can be probed effectively by stars.
    However, at larger distances planetary nebulae (PNe) replace stars as the tracer of the dynamics.
    Making use of the motion of PNe, \citet{Romanowsky03} measured the dynamics of three luminous elliptical galaxies (NGC821, NGC3379 and NGC4494)
    at large distances from the galactic center.
    They found that little dark matter is needed up to 6 effective radii.
    \citet{Milgrom&Sanders03} showed that this result can be understood in the framework of MOdified Newtonian Dynamics\,(MOND).
    As more data are available in the past decade, we revisit this problem.
    We combine PNe data (up to 6--8 effective radii) and stellar data from SAURON of 7 elliptical galaxies,
    including those 3 galaxies in \citet{Romanowsky03} with updated data and 4 other galaxies which have not been analyzed before.
    We conclude that the dynamics of these galaxies can be well explained by MOND.
    \end{abstract}


    \section{Introduction}\label{sec:intro}
    The so called ``missing mass problem'' has been a long standing issue in astrophysics.
    It was first mentioned by~\citet{Zwicky} that the gravitational mass inferred by virial theorem was larger than the luminous mass in Coma cluster.
    However, not until four decades later after the discovery of the flat rotation curve of spiral galaxies~\citep[e.g.,][]{Rubin},
    did scientists took this problem seriously.
    Now it ``appears'' everywhere: from galaxies, to cluster of galaxies, and even to cosmological scale phenomena
    such as large scale structure and comic microwave background radiation.

    In elliptical galaxies, the ``missing mass problem'' were found later by different methods such as gravitational lensing, x-ray emission of hot gas,
    and kinematics of matter.
    Gravitational lensing is a relativistic phenomenon predicted by General Relativity\,(GR), which has been tested successfully in the solar system.
    For large objects such as galaxies or clusters of galaxies, observations showed that the deflection angle of light ray is larger than the prediction of GR.
    This discrepancy is usually interpreted as dark matter in lens galaxies.
    Moreover, the Hubble constant $H_0$ obtained from gravitational time-delay in GR is larger than the value from supernova measurement~\citep{Riess11}.
    X-ray emitting hot gas can be used to estimate the gravitational potential of elliptical galaxies, in particular for those in the center of clusters.
    One can use stellar kinematics to estimate the mass of elliptical galaxies.
    However, stars are usually concentrated within one effective radius, $R_{\rm eff}$, and this makes the study of dynamics
    by stars at large distances very difficult.
    In order to study the dynamics of the galaxy up to several $R_{\rm eff}$, luminous objects such as, planetary nebulae\,(PNe), globular cluster,
    or satellite galaxies, are needed.

    \citet{Romanowsky03} reported the dynamics of elliptical galaxies by studying PNe up to 4--6 $R_{\rm eff}$.
    Surprisingly, little dark matter were found contained in the three luminous elliptical galaxies\,(NGC\,821, NGC\,3379, NGC\,4494)
    and the data is close to the model by Newtonian dynamics.
    The ``lack of dark matter'' can be explained by another view of the ``missing mass problem''.
    The crux of the matter is ``mismatch in acceleration''. This can be explained by unseen mass or modified theory of gravity.
    \citet{Milgrom&Sanders03} showed that those luminous elliptical galaxies reported by \citet{Romanowsky03}
    can be well explained by Modified Newtonian Dynamics\,(MOND).
    They pointed out that the systems belong to small acceleration discrepancy in MOND.

    With accumulating PNe data for elliptical galaxies in the past decade, we study the dynamics in 7 galaxies in the framework of MOND.
    The article is organized as follows.
    After a brief introduction of MOND in Section~2, we present the data of the elliptical galaxies in Section~3.
    We describe the mass model and dynamics in Section~4.
    Result and discussion are provided in Section~5.

    \section{MOND}\label{sec:MOND}

    The ``missing mass problem'' is, in fact, the mismatch between the measured gravitational acceleration and the inferred
    Newtonian gravitational acceleration (or GR) produced by the observed luminous matter of the system.
    The mismatch can be accounted for by the existence of non-luminous matter (a.k.a. dark matter), modified law of inertia or modified theory of gravity.
    \citet{Milgrom83} proposed MOND as a modification of Newton's second law when the acceleration is small than a small constant,
    $\mathfrak{a}_0\approx 1.2\times 10^{-10}$ m s$^{-2}$.
    It turned out that MOND can be expressed as a modified theory of gravity in the form of a nonlinear Poisson equation \citep[][]{BM84},
        \begin{equation}
        \label{eq:MOND}
          {\mathbf\nabla}\cdot\left[{\tilde\mu}(x){\mathbf g}\right]={\mathbf\nabla}\cdot{\mathbf g}_{\rm N}=-4\pi G\rho\,,
        \end{equation}
        where ${\mathbf g}$ is the gravitational acceleration in MOND, ${\mathbf g}_{\rm N}$ the gravitational acceleration in Newtonian gravity,
        and $x=|{\mathbf g}|/\mathfrak{a}_0$.
        ${\tilde\mu}(x)$ is called the interpolation function.
        It has the asymptotic behavior ${\tilde\mu}(x)\approx 1$ for $x\gg 1$ (Newtonian regime) and ${\tilde\mu(x)}\approx x$ for $x\ll 1$
        (deep MOND regime).
        \citet{ckt11} suggested a canonical interpolation function of the following form
        \begin{equation}
        \label{eq:canonicalmu}
          {\tilde\mu}(x)=\left[1-\frac{2}{(1+\eta x^\alpha)+\sqrt{(1-\eta x^\alpha)^2+4x^\alpha\,}}\right]^{1/\alpha}\,.
        \end{equation}

    From the kinematics of matter, MOND not only successfully explained the mass discrepancy (or the mismatch of acceleration) in spiral galaxies
    \citep{Sanders&McGaugh02}, but also the baryonic Tully-Fisher relation \citep{McGaugh11} in large acceleration mismatch systems.
    Recently, there are some studies on the kinematics of elliptical galaxies related to MOND, e.g., \citet{Rodrigues12}, \citet{Tortora14}.

    Relativistic effects such as gravitational lensing and time-delay can be studied under a relativistic gravity theory of MOND,
    e.g., T$e$V$e\,$S, GEA or BiMOND~\citep{TeVeS, Zlosnik07, Milgrom09}.
    \citet{ckt06} was the first to derive the gravitational lensing equation in relativistic MOND rigorously from T$e$V$e\,$S.
    This has been well tested on elliptical galaxy lens in quasar lensing \citep[see, e.g.,][]{Zhao06, ckt11} and also for quasar time-delay systems~\citep{Tian13}.
    In most quasar lensing cases, the system acceleration of the elliptical galaxy lenses is around $1\sim 10\,\mathfrak{a}_0$,
    and small acceleration discrepancy is expected.
    \citet{Sanders14} also tested strong lensing in MOND on the Sloan Lens Advanced Camera Surveys samples of elliptical lens to get
    a luminous mass consistent with stellar mass determined from population synthesis models using Salpeter initial mass function (IMF).

    Hot gas in elliptical galaxies emit x-ray.
    \citet{Milgrom12} found that MOND fitted well with the data from NGC\,720 and NGC\,1521 to very large galactic radii ($100$ and $200$ kpc),
    which correspond to a wide range of acceleration from $0.1\,\mathfrak{a}_0$ to more than $10\,\mathfrak{a}_0$.
    In addition, globular clusters can be used to study the dynamics of elliptical galaxies outside several effective radii in
    MOND~\citep[see, e.g.,][]{Richtler08, Samurovic&Cirkovic08, Schuberth12, Samurovic12, Samurovic14}.
    \citet{Samurovic14} pointed out that although MOND is successful in explaining the dynamics of less massive fast rotators,
    it cannot fit several massive slow rotators well without an additional dark matter.

    A convenient parameter
    \begin{equation}
    \label{eq:xi}
      \xi\equiv\left(\frac{GM}{R_{\rm eff}^2\,\mathfrak{a}_0}\right)^{1/2}\,,
    \end{equation}
    has been used to distinguish between different regimes, say, Newtonian and deep MOND regimes \citep{Milgrom83b,Milgrom&Sanders03}.
    For spiral galaxies in deep MOND regime, the terminal velocity is given by
    $V_\infty^2=\sqrt{GM\mathfrak{a}_0}$, and Equation~(\ref{eq:xi}) can be expressed as
    $\xi=V_{\infty}^2/(R_{\rm eff}\,\mathfrak{a}_0)$ \citep{Milgrom83b}.
    Equivalently, instead of Equation~(\ref{eq:xi}),
    one can define a critical surface density $\Sigma_{\rm c}=\mathfrak{a}_0/G$~\citep{Milgrom&Sanders03}.

    In the case of spiral galaxies, $\xi>1$ and $\xi<1$ (or $\Sigma>\Sigma_{\rm c}$ and $\Sigma<\Sigma_{\rm c}$)
    correspond to high surface brightness (HSB) spirals and low surface brightness (LSB) spirals, respectively.
    In the framework of dark matter, HSB spirals require less dark matter while LHS spirals more.
    In the framework of MOND, $\xi$ tells us the whether the system is more like a Newtonian system (HSB when $\xi>1$) or a MONDian system (LSB when $\xi<1$).

    In the case of elliptical galaxies, let us take the three galaxies in \citet{Romanowsky03} as examples.
    \citet{Milgrom&Sanders03} estimated luminous the masses by assuming mass-to-light
    ratio equal to 4, and
    got $\xi\approx\,5.7$ for NGC\,3379, $\xi\approx\,3.6$ for NGC\,821 and $\xi\approx\,3.4$ for NGC\,4494.

    \section{Data}\label{sec:Data}

            \begin{table*}
             \centering
             \setlength{\extrarowheight}{3pt}
            \caption[\textbf{The sample of elliptical galaxies.}]
            {\textbf{The sample of elliptical galaxies.}}\label{tab:data}
            \begin{tabular}{ccccccccc}
            \hline
            Name & Type & D   &     V$_{\rm los}$       & M$_{\rm B}$ & N$_{\rm PNe}$ & R$_{\rm eff}$ & R$_{\rm last}$ & References \\
             &      & Mpc & km\,s$^{-1}$ & mag     &               & arcsec             &      arcmin         &            \\
            (1) & (2) & (3) & (4) & (5) & (6) & (7) & (8) & (9) \\
            \hline
            NGC\,821     &   E6  &   23.4    &   1718    &   -20.81   &   127 &   39.8  &    6.8    & \citet{Coccato09}\\
            NGC\,1344   &   E5  &   18.4    &   1169    &   -19.66   &   194 &   46  &    6.7 & \citet{Coccato09}\\
            NGC\,3379    &   E1  &   10.3 &  918  &   -20.67   &   214 &  39.8  &   7.2         & \citet{Douglas07}\\
            NGC\,4374    &   E1  &   18.5    &   1017    &   -21.21   &   454 &   52.5  &    6.9    & \citet{Coccato09}\\
            NGC\,4494    &   E1  &   16.6    &   1342    &   -21.12   &   267 &   50  &    7.6    & \citet{Napolitano09}\\
            NGC\,4697   &   E6  &   11.4    &   1252    &   -21.22   &   535 &   61.7  &    6.6    & \citet{Coccato09}\\
            NGC\,5846    &   E0  &   24.2    &   1712    &   -21.33   &   124 &   58.9  &    6  & \citet{Coccato09}\\
            \hline
            \end{tabular}
            \begin{minipage}{15.5cm}
            \textit{Notes.}
            (1) name of galaxy,
            (2) morphological type of galaxy in NED,
            (3) distance from ATLAS$^{\rm 3D}$ \citep{Cappellari11},
            (4) galaxy heliocentric systemic velocity in NED,
            (5) absolute B-band  magnitude from HyperLeda database \citep{Makarov09},
            (6) total number of PNe with measured radial velocities from \citet{Coccato09},
            (7) effective radius from ATLAS$^{\rm 3D}$ database \citep{Cappellari11},
            (8) maximum distance of the PNe detections from the galaxy center \citep{Coccato09},
            (9) reference of the PNe data.
            \end{minipage}
            \end{table*}

        \begin{figure}
            \centering
            \includegraphics[width=\columnwidth]{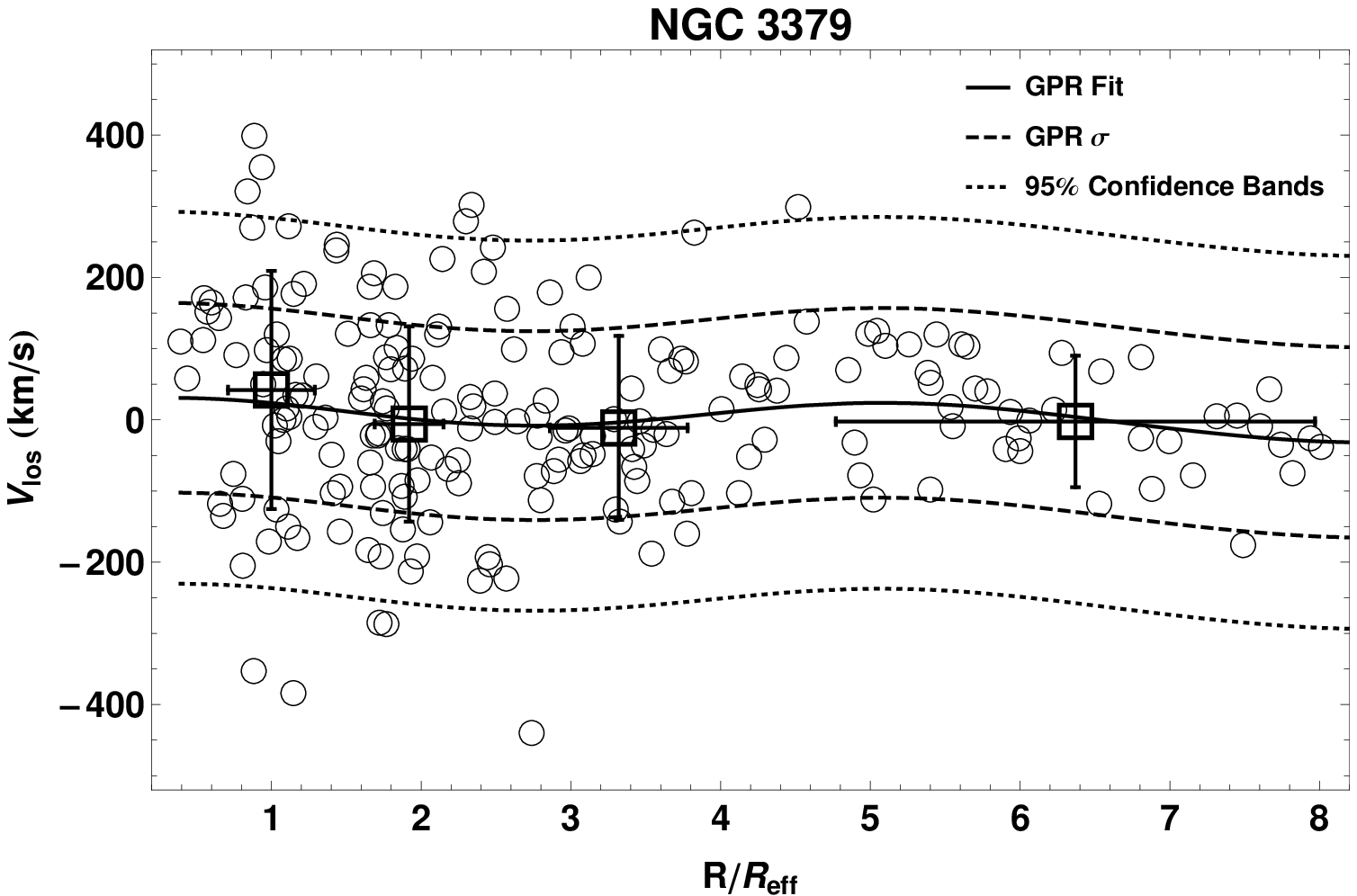}
            \includegraphics[width=\columnwidth]{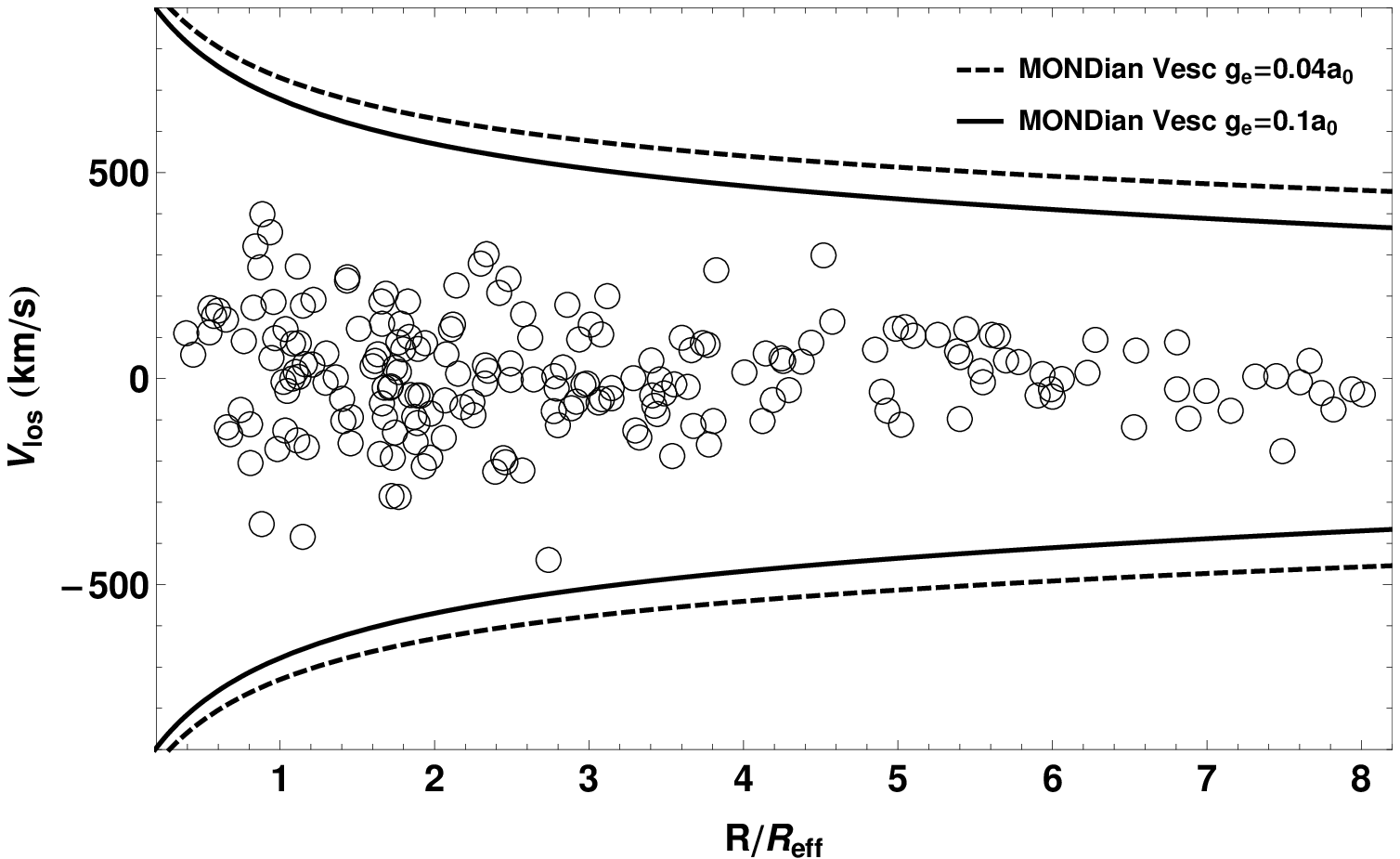}
            \caption{
            Upper panel: Line-of-sight velocity of 214 PNe in NGC\,3379 (circles) is plotted against their distance from the galactic center
            (data from PN.S public website).
            The four squares are the average line-of-sight velocity from maximum likelihood, and the vertical error bar represents
            the line-of-sight velocity dispersion.
            The solid line is the mean velocity obtained by Gaussian Process Regression\,(GPR).
            The dash line and the dotted line are the error and 95\% confidence band by GPR, respectively.
            Lower panel: escape speed of a test particle in NGC\,3379.
            Due to the external field effect, particles can escape a MONDian system if the system is embedded in a constant external field \citep[see, e.g.,][]{BM84,Famaey07}.
            The solid line is the escape speed for an external field $0.1\mathfrak{a}_0$ and dash line for $0.04\mathfrak{a}_0$.
            Simple form is used in this example.
            }
            \label{fig:PNe}
        \end{figure}

    As the surface brightness from the stars of elliptical galaxies drops rapidly beyond 2 $R_{\rm eff}$, the kinematics of the galaxy at
    larger distances has to be studied by other sources.
    Motion of gas can be used in spiral galaxies but this is not suitable for elliptical galaxies as their gas content is low.
    Instead, planetary nebulae\,(PNe) and globular clusters have been suggested for elliptical galaxies.
    PNe are good tracers for velocity distribution because of their strong emission line O[III] at 500.7 nm.

    The first PN radial velocity measurement was made in the halo of Centaurus A at the Anglo-Australian Telescope \citep{Hui95} by using
    a multi-fiber instrument.
    The success of the project led to a special-purpose instrument, the Planetary Nebulae Spectrograph\,(PN.S)
    mounted on the William Herschel Telescop in La Palma \citep{Douglas97}.

    The PN.S \citep{Douglas02} split the incoming light into two separate cameras and was able to measure simultaneously the position,
    radial velocity and flux of the PNe in the target galaxies.
    Since its first-light in 2001, PN.S focused on elliptical S0 galaxies.
    Its data is publicly available in PN.S public website\footnote{$http://www.strw.leidenuniv.nl/pns/PNS\_public\_web$}.
    Its catalogue lists positions, observed wavelength, heliocentric radial velocity, and magnitude of PNe~\citep{Douglas07, Coccato09, Napolitano09}.
    The PNe line-of-sight velocity distribution of NGC\,3379
    is shown in the upper panel of
    Figure~\ref{fig:PNe}.
    The velocity dispersion of NGC\,3379 is obtained from the distribution by maximum likelihood method for 4 bins.
    This is consistent with the analysis by Gaussian Process Regression, see
    upper panel of Figure~\ref{fig:PNe}.
    We also obtain the velocity dispersion of NGC\,4494 and NGC\,821 by the same procedure.
    For NGC\,5846, NGC\,4374, NGC\,1344 and NGC\,4697 we take the values published in SAURON~\citep{Coccato09}.

    For illustration we plot the escape speed of a test particle in NGC\,3379.
    Recall that the MONDian potential at large distance from a galaxy (a finite object) is $\sim \log r$, thus every test particle is bound.
    However, if the galaxy is embedded in a constant external field ${\bf g}_e$,
    the test particle can escape due to the external field effect in MOND \citep[see, e.g.,][]{BM84,Famaey07,FM12}.
    Adopting the procedure of \citet{Famaey07}, and choosing simple form
    ($\mu(x)=x/(1+x)$), we have
    \citep[cf. Equation~(5) of][]{Famaey07}
    \begin{equation}
    \frac{g(g+g_e)}{g+g_e+\mathfrak{a}_0}=\frac{Gm(r)}{r^2}=g_{\rm N}\,,
    \end{equation}
    where $g=|{\bf g}|$, $g_e=|{\bf g}_e|$, and $m(r)$ is the mass enclosed within $r$.
    The escape speed can be computed from
    \begin{equation}
    v_{esc}(R)=\left[2\int_R^{\infty}g(r)\,dr\right]^{1/2}\,.
    \end{equation}
    We adopt Hernquist model for the mass distribution, i.e., $m(r)=Mr^2/(r+r_h)^2$, where we use the Salpeter mass for $M$ \citep{Cappellari13a,Cappellari13b}.
    The lower panel of Figure~\ref{fig:PNe} shows the escape speed of two cases: $g_e=0.1\mathfrak{a}_0$ and $0.04\mathfrak{a}_0$.

    For the stellar component, SAURON integral-field unit on the William Herschel Telescope has produced excellent
    two dimensional velocity dispersion data of nearby elliptical galaxies~\citep{Coccato09}.
    By folding the major axis and the minor axis components \citep[e.g.,][]{Romanowsky03}, we combine the data from stellar component with data
    from PNe component to study the dynamics of the sample galaxies up to several $R_{\rm eff}$.

    There are 16 galaxies with publicly available PNe data posted in the PN.S public website.
    Among these, 8 are classified as S0 and 8 elliptical galaxies.
    In this work, we focus on elliptical galaxies only.
    The elliptical galaxy NGC\,4283 has only 11 PNe data which is not enough for our analysis.
    Thus we have 7 elliptical galaxies in our sample: NGC\,821, NGC\,1344, NGC\,3379, NGC\,4374, NGC\,4494, NGC\,4697, and NGC\,5846.
    Table~\ref{tab:data} lists some properties of these galaxies.

    \section{Model}\label{sec:Mod}
        For simplicity, we model an elliptical galaxy as a spherically symmetric stellar system.

    \subsection{Velocity dispersion}\label{sec:Jeans}

        The velocity dispersion a spherically symmetric stellar system in equilibrium is governed by the Jeans equation in spherical coordinates
        \citep[see, e.g.,][]{Binney2008},
        \begin{equation}
        \label{eq:Jeans}
        \frac{d(\rho\sigma_r^2)}{dr}+\frac{2\beta}{r}\rho\sigma_r^2=-\rho g\,,
        \end{equation}
        where $\beta=1-(\sigma_t^2/\sigma_r^2)$ is the anisotropy parameter\,($\beta=0$ for the isotropy case).

        The velocity dispersion measured along the line-of-sight at projected radius $R$ is given by
        \begin{equation}
        \label{eq:vel_I}
        \sigma_I^2(R)=\frac{2}{I(R)}\int_{R}^{\infty}\,\sigma_r^2\left(1-\beta(r)\frac{R^2}{r^2}\right)\frac{\rho(r)rdr}{\sqrt{r^2-R^2}}\,,
        \end{equation}
        where the surface density is
        \begin{equation}
        \label{eq:surface}
        I(R)=2\int_{R}^{\infty}\,\frac{\rho(r') r'\,dr'}{\sqrt{r'^2-R^2}}\,.
        \end{equation}

        In this paper we consider isotropic model (i.e., $\beta=0$) and a particular anisotropic model
        \begin{equation}
        \label{eq:anisotropic}
        \beta=\frac{r^2}{r_{\rm a}^2+r^2}\,.
        \end{equation}
        This anisotropic model can be formed by dissipationless collapse systems \citep[see, e.g.,][]{vanAlbada82,Milgrom&Sanders03}.

    \subsection{Mass model}\label{sec:Hernquist}

    \begin{figure}
            \centering
            \includegraphics[width=\columnwidth]{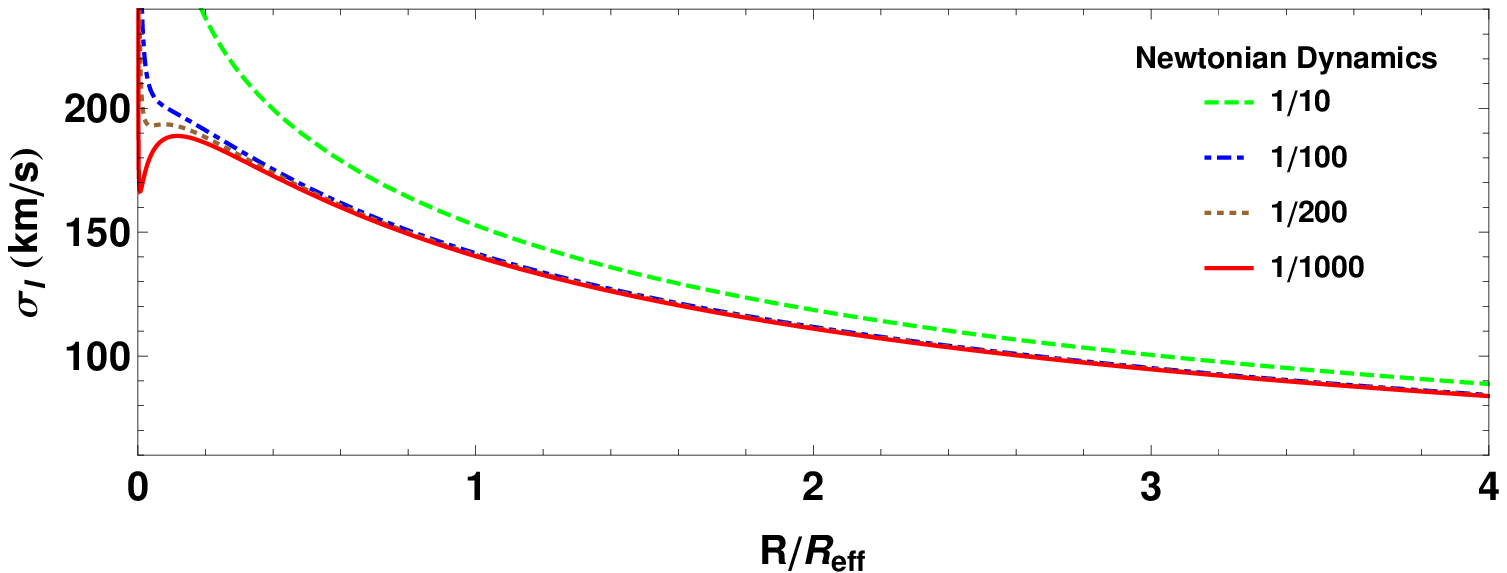}
            \includegraphics[width=\columnwidth]{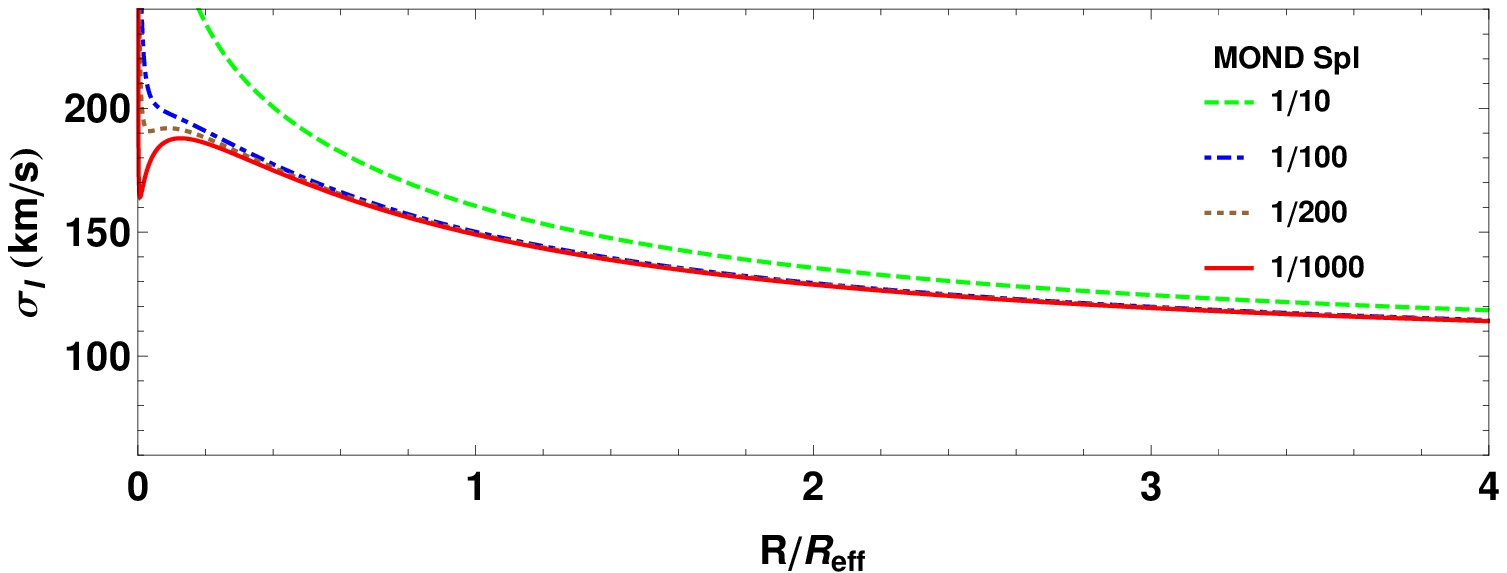}
            \caption{
            The projected line-of-sight velocity dispersion profile of a galaxy.
            The mass distribution of the galaxy is described by Hernquist model.
            Newtonian gravity and simple form in MOND are used as the examples.
            Besides the self-gravity of the mass distribution, the influence of a SMBH at the center is considered.
            The dash line represents SMBH to galaxy mass ratio at 1/10, dotted-dash line at 1/100, dotted line at 1/200, and solid line at 1/1000.
            (Color in online version: dash, dotted-dash, dotted and solid correspond to green, blue, orange and red, respectively.)
            }
            \label{fig:BH}
        \end{figure}

        We adopt Hernquist profile~\citep{Hernquist90} as the mass model for the elliptical galaxies,
        \begin{equation}
            \label{eq:Hernquist}
            \rho_{\rm h}(r)=\frac{M_{\rm h}\,r_{\rm h}}{2\pi r(r+r_{\rm h})^3}\,,
            \quad
            g_{\rm h}(r)=\frac{G\,M_{\rm h}}{(r+r_{\rm h})^2}\,,
        \end{equation}
        where $M_{\rm h}$ is the mass of the galaxy and $g_{\rm h}$ is the corresponding Newtonian gravitational acceleration.
        We note again that we take the convention that positive gravitational acceleration means towards the origin.
        Assuming constant mass-to-light ratio, the surface brightness distribution is the same as the surface density distribution,
        and the effective radius $R_{\rm eff}$ is the same as half-mass radius.
        In this case, $r_h\approx0.551\,R_{\rm eff}$ and this is the relation we use in our modeling.

        As stellar data in the innermost region indicates the need of a supermassive black hole (SMBH), we also include a point mass
        at the center to represent the SMBH
        \begin{equation}
        \label{eq:blackhole}
          g_{\rm bh}(r)=\frac{GM_{\rm bh}}{r^2}\,.
        \end{equation}
        Figure~\ref{fig:BH} shows the effect of a SMBH in a Hernquist model on velocity dispersion.
        Different curves in the figure represent different SMBH mass to galaxy mass ratio ($M_{\rm bh}/M_{\rm h}$).
        In our sample, this ratio is within the range 1/700 -- 1/200.
        For mass ratio less than 1/100 ($M_{\rm bh}/M_{\rm h}<1/100$), the effect of the SMBH can be seen within 0.2 $R_{\rm eff}$ only.

        We use simple singular isothermal model for the dark matter halo when we compared MOND and Newtonian gravity of the galaxies later,
        \begin{equation}
        \label{eq:isothermal}
          g_{\rm iso}(r)=\frac{2\sigma_v^2}{r}\,.
        \end{equation}

    \subsection{Gravity model}\label{sec:MONDgravity}
        We adopt MOND for the gravitational acceleration in Section~\ref{sec:Jeans}.
        For spherically symmetric systems, Equation~(\ref{eq:MOND}) can be ``inverted'' to
        \begin{equation}
        \label{eq:sphericalMOND}
          {\bf g}=-\,g\,{\hat{\bf e}}_r={\tilde\nu}(x_{\rm N})\,{\bf g}_{\rm N}=-\,{\tilde\nu}(x_{\rm N})\,g_{\rm N}\,{\hat{\bf e}}_r\,,
        \end{equation}
        where $x_{\rm N}=|{\bf g}_{\rm N}|/\mathfrak{a}_0=g_{\rm N}/\mathfrak{a}_0$
        and ${\tilde\nu}(x_{\rm N})$ is called the inverted interpolation function.
        The canonical form of this function can be obtained from Equation~(\ref{eq:canonicalmu}) \citep[see,][]{ckt11},
        \begin{equation}
        \label{eq:canonicalnu}
          \tilde{\nu}(x_{\rm N})=\left[1+\frac{1}{2}\left(\sqrt{4x_{\rm N}^{-\alpha}+\eta^2}-\eta\right)\right]^{1/\alpha}\,.
        \end{equation}
        For instance, in this paper we use
        (1) $(\alpha,\eta)=(2,1)$, where Equation~(\ref{eq:canonicalmu}) becomes $\tilde{\mu}(x)=x/\sqrt{1+x^2}$
        and is called the standard form \citep[e.g.,][]{Milgrom83}, and
        (2) $(\alpha,\eta)=(1,1)$, where $\tilde{\mu}(x)=x/(1+x)$ and is called the simple form \citep[e.g.,][]{FB05}.

        To compute the velocity dispersion $\sigma_I$ in Equation~(\ref{eq:vel_I}),
        we adopt the mass distribution $\rho_{\rm h}$ of the Hernquist model in Equation~(\ref{eq:Hernquist}).
        In addition to the self-gravity of the mass distribution, the gravitational acceleration on the system
        may consist of a contribution from a central SMBH.
        To compute the gravitational acceleration in MOND we set $g_{\rm N}=g_{\rm h}+g_{\rm bh}$ and then
        use Equation~(\ref{eq:sphericalMOND}) to get $g$ for Equation~(\ref{eq:vel_I}).
        For comparison, we also compute the Newtonian gravitational acceleration with a dark matter halo,
        i.e., $g=g_{\rm N}=g_{\rm h}+g_{\rm iso}$.

    \section{Result and discussion}\label{sec:res}

    \begin{figure*}
            \centering
            \includegraphics[width=\columnwidth]{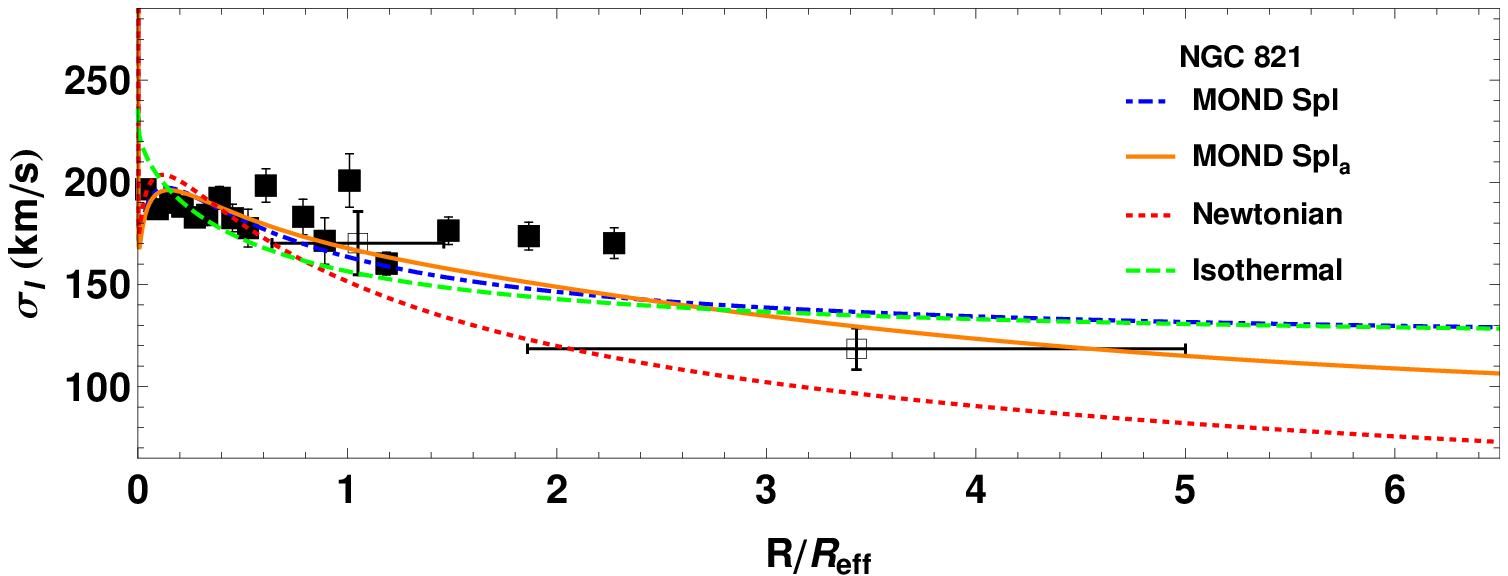}
            \includegraphics[width=\columnwidth]{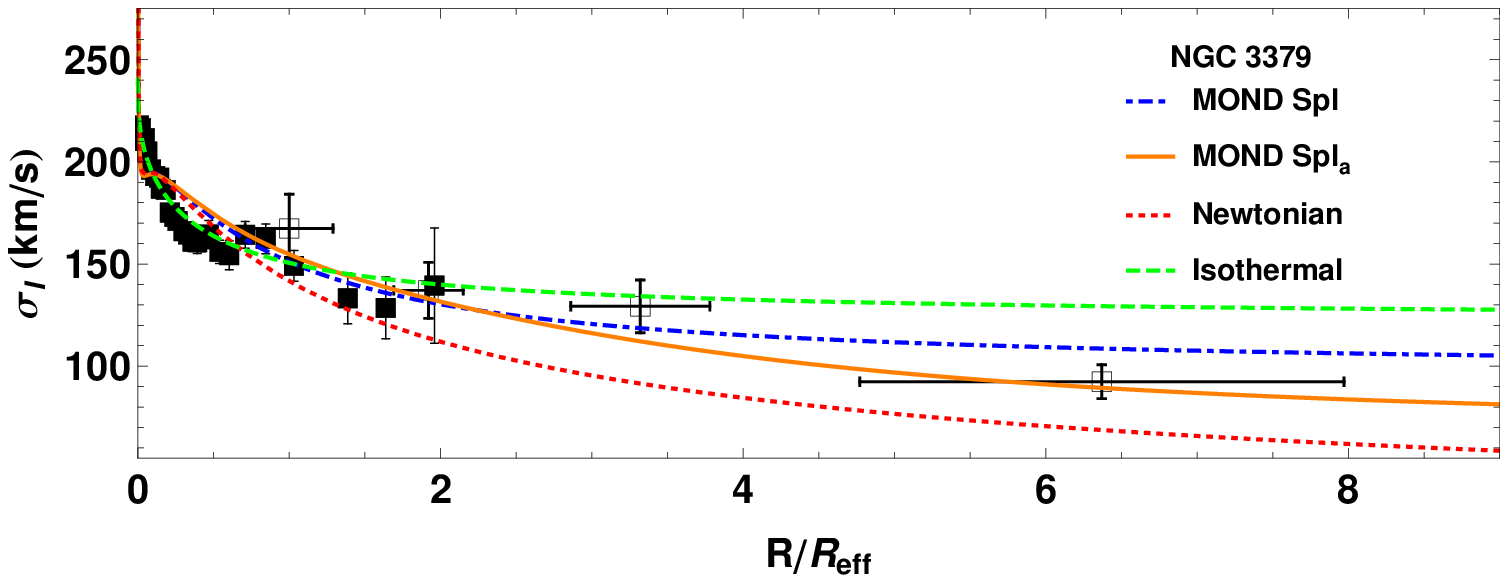}
            \includegraphics[width=\columnwidth]{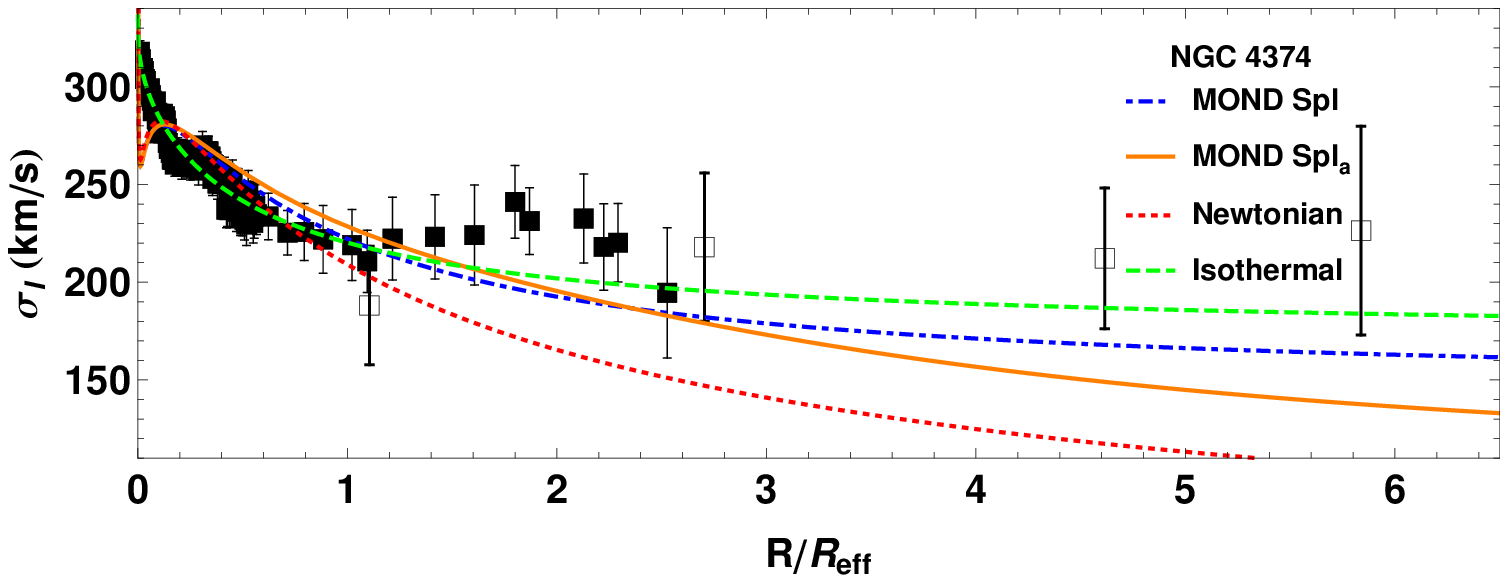}
            \includegraphics[width=\columnwidth]{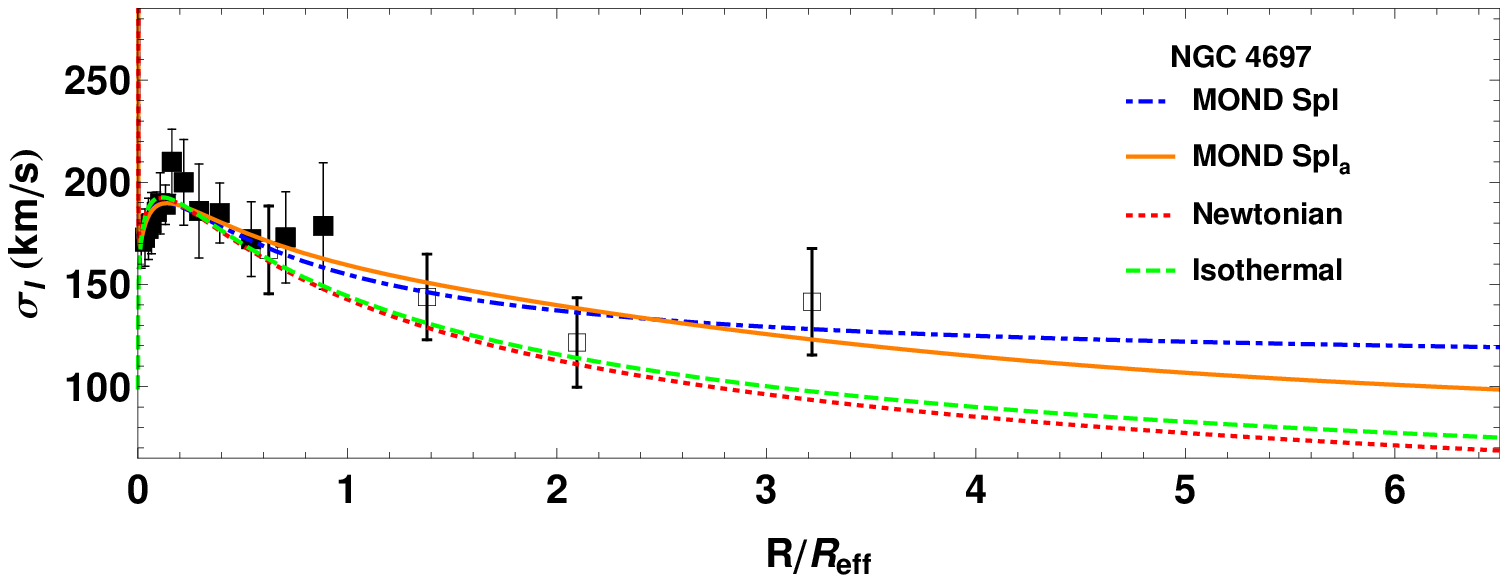}
            \caption{
            The projected line-of-sight velocity dispersion of NGC\,4697, NGC\,821, NGC\,3379 and NGC\,4374 (see Table~\ref{tab:data})
            as a function of the projected radius normalized to $R_{\rm eff}$ of individual galaxy.
            Open circles are PNe data and solid circles are stellar data.
            Dotted-dash line and solid line correspond to MOND, and dashed line and dotted line to Newtonian gravity.
            Dotted-dash line is the best fit of Hernquist isotropic model with SMBH in MOND in simple form in Equation~(\ref{eq:canonicalnu}).
            Solid line is the same as dotted-dash line except that it is an anisotropic model with $r_{\rm a}=3 R_{\rm eff}$ (see Equation~(\ref{eq:anisotropic})).
            Dotted line is Hernquist model with SMBH (but without dark matter halo) in Newtonian gravity.
            Dash line is the best fit of Hernquist model with isothermal dark matter halo in Newtonian gravity.
            The masses of the SMBH for the solid line, dotted-dash line and the dotted line are prescribed and their values are
            adopted from \citet{McConnell11}.
            Each of these three lines is the result of one-parameter fitting, the mass of the galaxy $M_{\rm h}$, while the dash line is
            the result of two-parameter fitting, the mass of the galaxy $M_{\rm h}$ and the parameter of the singular isothermal sphere $\sigma_v$,
            see Table~\ref{tab:parameterfix}.
            (Color in online version: dotted-dash, solid, dotted and dash correspond to blue, red, orange and green, respectively.)
            }
            \label{fig:fix}
        \end{figure*}

        \begin{table*}
            \centering
            \tabcolsep=3pt
            \setlength{\extrarowheight}{5pt}
            \caption[\textbf{The fitting parameters of Figure~\ref{fig:fix}.}]
            {\textbf{The fitting parameters of Figure~\ref{fig:fix}.}}\label{tab:parameterfix}
            \begin{tabular}{ccccccccc}
            \hline
            Name & $M_{\rm h}$ & $M_{\rm dm(<5R_{\rm eff})}$ & $M_{\rm h}$ & $M_{\rm h}$ & $M_{\rm h}$ & $M_{\rm bh}$ & $M_{\rm Sal}$ & $\xi$ \\
                 & Iso & Iso & New & Spl$_{a}$ & Spl &  & Sal & \\
            (1) & (2) & (3) & (4) & (5) & (6) & (7) & (8) & (9) \\
            \hline
            NGC\,821 & 7.0$\pm1.7$ & 29.8$^{+3.16}_{-3.0}$ & 23.3$\pm0.34$ & 19.6$\pm0.32$ & 20.3$\pm0.33$ & $1.8^{+0.8}_{-0.8}$ & 11.1 & 2.7\\
            NGC\,3379 & 1.8$\pm0.28$ & 13.9$^{+0.46}_{-0.45}$ & 8.9$\pm0.08$ & 8.4$\pm0.07$ & 8.6$\pm0.08$ & $4.6^{+1.1}_{-1.2}$ & 11.1 & 5.1 \\
            NGC\,4374 & 13.2$\pm0.90$ & 63.8$^{+1.64}_{-1.61}$ & 46.2$\pm0.21$ & 43.1$\pm0.20$ & 43.7$\pm0.21$ & $8.5^{+0.9}_{-0.8}$ & 41.1 & 5.0 \\
            NGC\,4697 & 15.0$\pm2.92$ & 1.67$^{+8.22}_{-1.35}$ & 15.6$\pm0.66$ & 14.0$\pm0.64$ & 14.3$\pm0.65$ & $2.0^{+0.2}_{-0.2}$ & 14.6 & 3.7 \\
            \hline
            \end{tabular}
            \begin{minipage}{15.5cm}
            \textit{Notes.}
            (1) name of galaxy,
            (2) fitting luminous mass of the galaxy,
            (3) fitting mass of dark matter halo within 5 effective radii,
            (4) fitting mass of the galaxy in Newtonian gravity,
            (5) fitting mass of the galaxy in MONDian gravity in simple form with anisotropic parameter $\beta$,
            (6) fitting mass of the galaxy in MONDian gravity in simple form,
            (7) mass of the SMBH (in $10^{8}$ $M_{\odot}$) from \citet{McConnell11},
            (8) mass estimated from population synthesis models with Salpeter IMF \citep[ATLAS$^{\rm 3D}$,][]{Cappellari13a, Cappellari13b},
            (9) the parameter $\xi$ in Equation~(\ref{eq:xi}) with mass given by column (8).
            All masses are in unit of $10^{10}$ $M_{\odot}$ (except mass of SMBH).
            Columns (2) \& (3) are the results of two-parameter fitting in Newtonian gravity
            (luminous mass $M_{\rm h}$ and parameter of the isothermal dark matter halo $\sigma_v$).
            Columns (4), (5) \& (6) are the results of one-parameter fitting (galaxy mass $M_{\rm h}$ with the mass of the SMBH given by Column (7)).
            \end{minipage}
        \end{table*}

        \begin{figure*}
            \centering
            \includegraphics[width=\columnwidth]{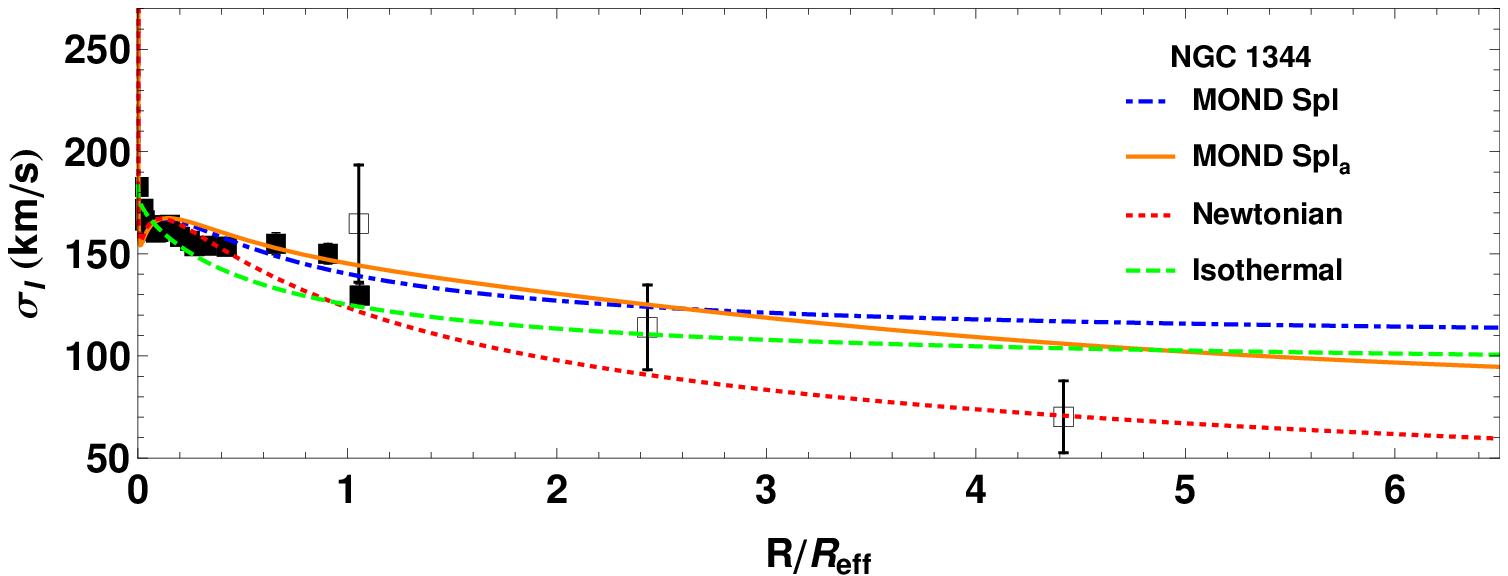}
            \includegraphics[width=\columnwidth]{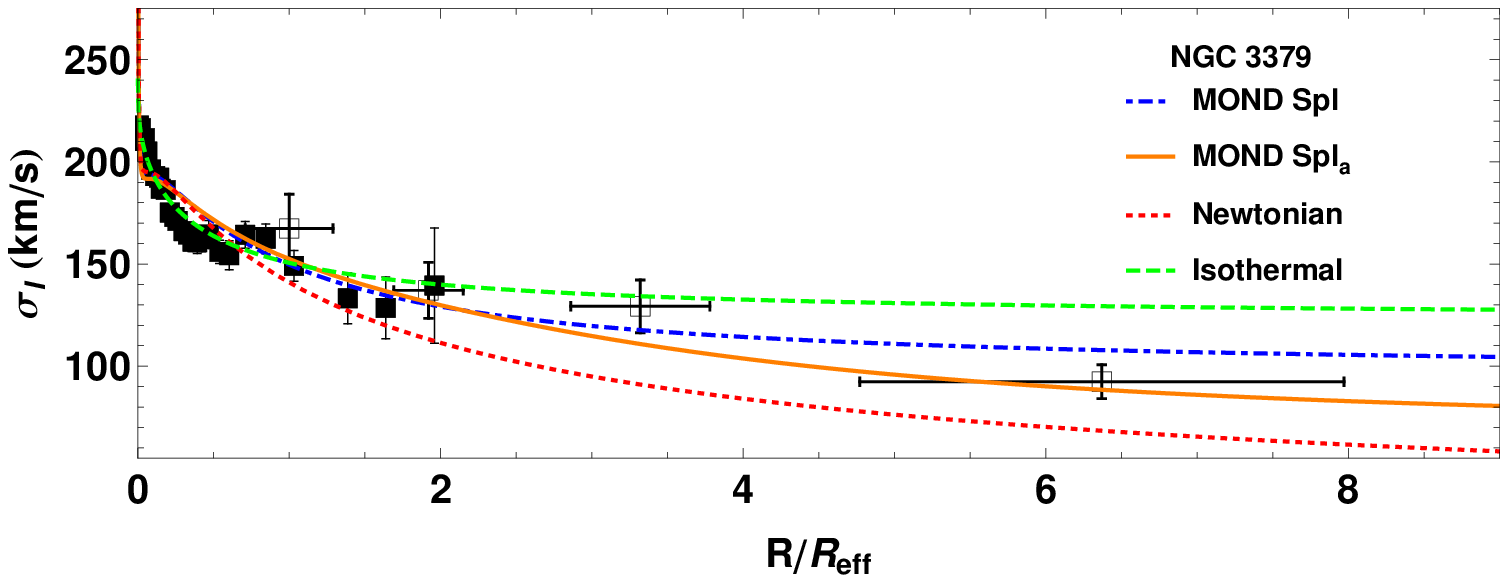}
            \includegraphics[width=\columnwidth]{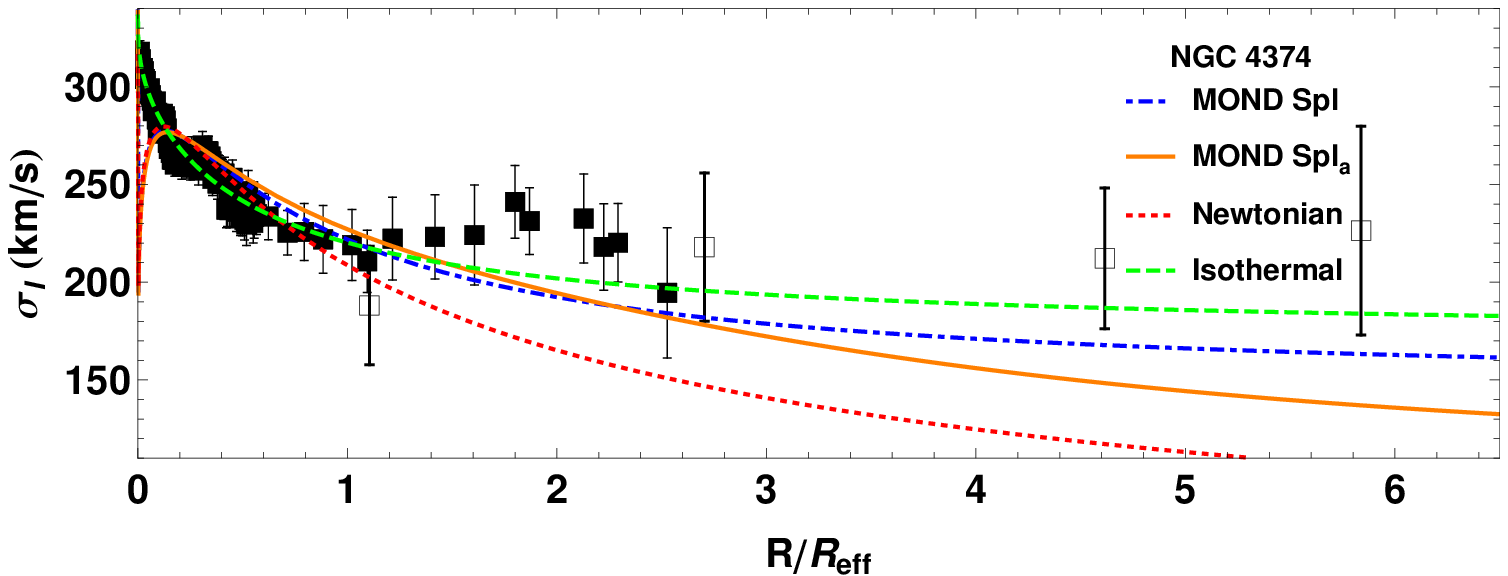}
            \includegraphics[width=\columnwidth]{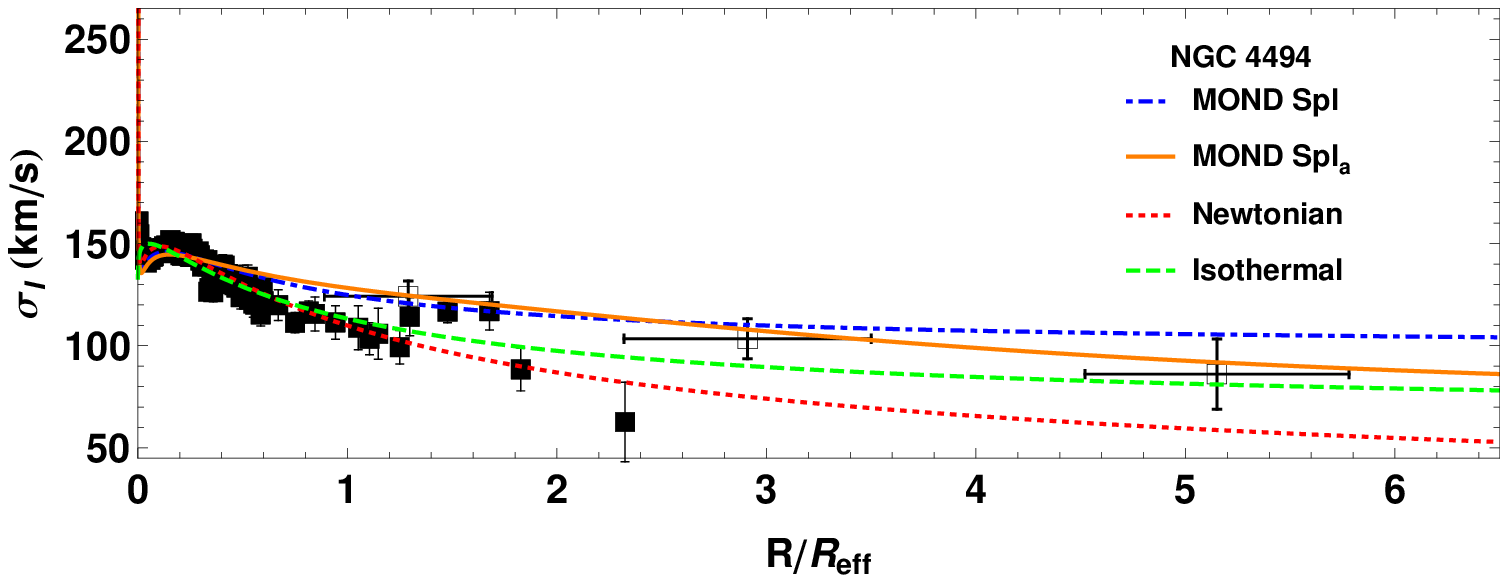}
            \includegraphics[width=\columnwidth]{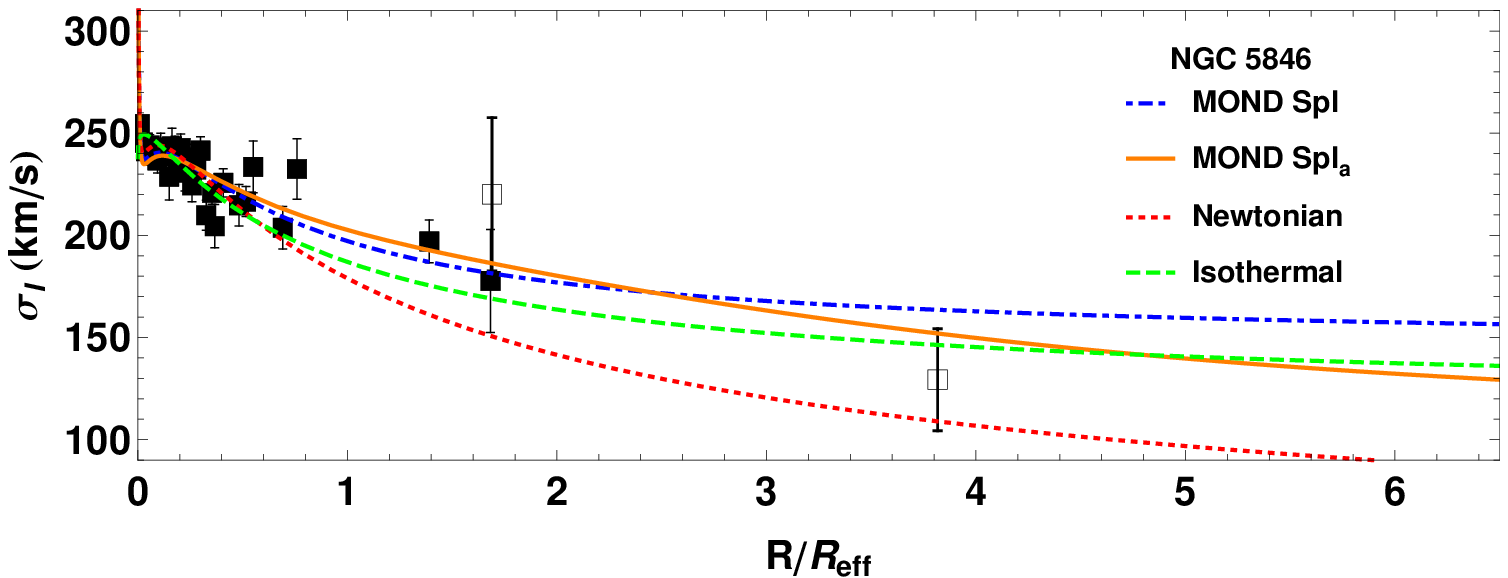}
            \caption{
            The same as Figure~\ref{fig:fix} but for the galaxies NGC\,1344, NGC\,3379, NGC\,4374, NGC\,4494, and NGC\,5846.
            All the lines are the result of two-parameter fitting.
            The fitting parameters for dotted-dash line, solid line, and dotted line are the mass of the galaxy and SMBH ($M_{\rm h}$ and $M_{\rm bh}$),
            see Table~\ref{tab:parametertwo}.
            Dash line is the same as in Figure~\ref{fig:fix}, i.e., $M_{\rm h}$ and $\sigma_v$.
            (Color in online version: dotted-dash, solid, dotted and dash correspond to blue, red, orange and green, respectively.)
            }
            \label{fig:two}
        \end{figure*}

        \begin{table*}
            \centering
            \tabcolsep=1pt
            \setlength{\extrarowheight}{5pt}
            \caption[\textbf{The fitting parameters of Figure~\ref{fig:two}.}]
            {\textbf{The fitting parameters of Figure~\ref{fig:two}.}}\label{tab:parametertwo}
            \begin{tabular}{cccccccccccc}
            \hline
            Name & $M_{\rm h}$ & $M_{\rm dm(<5R_{\rm eff})}$ & $M_{\rm h}$ & $M_{\rm bh}$ & $M_{\rm h}$ & $M_{\rm bh}$ & $M_{\rm h}$ & $M_{\rm bh}$ & $M_{\rm bh}$
            & $M_{\rm Sal}$ & $\xi$ \\
                  & Iso & Iso & New & New & Spl$_{a}$ & Spl$_{a}$ & Spl & Spl &  & Sal & \\
            (1) & (2) & (3) & (4) & (5) & (6) & (7) & (8) & (9) & (10) & (11) & (12) \\
            \hline
            NGC\,1344 & 4.7$\pm0.22$ & 16.3$^{+0.31}_{-0.31}$ & 14.1$\pm0.5$ & 2.68$\pm0.1$ & 12.6$\pm0.5$ & 2.77$\pm0.1$ & 12.8$\pm0.5$ & 2.81$\pm0.1$ & $-$ & (16.8) & (5.3) \\
            NGC\,3379 & 1.8$\pm0.28$ & 13.9$^{+0.46}_{-0.45}$ & 8.8$\pm0.6$ & 5.19$\pm0.3$ & 8.1$\pm0.6$ & 5.59$\pm0.4$ & 8.4$\pm0.6$ & 5.38$\pm0.4$ & $4.6^{+1.1}_{-1.2}$ & 11.1 & 5.1 \\
            NGC\,4374 &  13.2$\pm0.90$ & 63.8$^{+1.64}_{-1.61}$ & 46.6$\pm2.1$ & 11.1$\pm0.5$ & 42.6$\pm2.1$ & 12.4$\pm0.6$ & 43.7$\pm2.1$ & 11.8$\pm0.6$ & $8.5^{+0.9}_{-0.8}$ & 41.1 & 5.0\\
            NGC\,4494 & 7.0$\pm0.21$ & 7.7$^{+0.37}_{-0.36}$ & 10.9$\pm0.3$ & 2.29$\pm0.1$ & 8.9$\pm0.3$ & 2.85$\pm0.1$ & 9.3$\pm0.3$ & 2.7$\pm0.1$ & $-$ & 12.4 & 3.0 \\
            NGC\,5846 & 28.0$\pm2.08$ & 44.0$^{+3.54}_{-3.40}$ & 49.5$\pm7.6$ & 18.9$\pm2.9$ & 42.9$\pm7.6$ & 22.3$\pm3.9$ & 44.3$\pm7.7$ & 21.3$\pm3.7$ & $-$ & 32.9 & 3.3 \\
            \hline
            \end{tabular}
            \begin{minipage}{15.5cm}
            \textit{Notes.}
            (1) name of galaxy,
            (2) fitting luminous mass of the galaxy,
            (3) fitting mass of dark matter halo within 5 effective radii,
            (4) fitting mass of the galaxy in Newtonian gravity,
            (5) fitting SMBH mass in Newtonian gravity,
            (6) fitting mass of the galaxy in MONDian gravity in simple form with anisotropic parameter $\beta$,
            (7) fitting SMBH mass in MONDian gravity in simple form with anisotropic parameter $\beta$,
            (8) fitting mass of the galaxy in MONDian gravity in simple form,
            (9) fitting SMBH mass in MONDian gravity in simple form,
            (10) mass of the SMBH from \citet{McConnell11},
            (11) mass estimated from population synthesis models with Salpeter IMF \citep[ATLAS$^{\rm 3D}$,][]{Cappellari13a, Cappellari13b},
            (12) the parameter $\xi$ of Equation~(\ref{eq:xi}) with mass given by column (11).
            $M_{\rm h}$ and $M_{\rm dm}$ are in unit of $10^{10}$ $M_{\odot}$ while $M_{\rm bh}$ in $10^8$ $M_\odot$.
            Columns (2) \& (3) are the results of two-parameter fitting in Newtonian gravity
            (luminous mass $M_{\rm h}$ and parameter of the isothermal dark matter halo $\sigma_v$).
            Columns (4) \& (5) are the results of two-parameter fitting in Newtonian gravity
            (galaxy mass $M_{\rm h}$ and SMBH mass $M_{\rm bh}$).
            (6) \& (7) and (8) \& (9) are the same as (4) \& (5) except they are the results of
            MONDian gravity in standard form and simple form, respectively.
            We note that in the case of NGC\,1344, we do not have the mass from population synthesis.
            Its mass in Column (11) is estimated from its luminosity $4.2\times 10^{10}$ $L_\odot$ \citep{Teodorescu05}
            with a mass-to-light ratio equal to 4 \citep{Milgrom&Sanders03}.
            \end{minipage}
        \end{table*}

        \begin{table*}
            \centering
            \setlength{\extrarowheight}{5pt}
            \caption[\textbf{The fitting parameters for Figure~\ref{fig:fix} \& \ref{fig:two} in term of mass-to-light ratio.}]
            {\textbf{The fitting parameters for Figure~\ref{fig:fix} \& \ref{fig:two} in term of mass-to-light ratio.}}\label{tab:masstolight}
            \begin{tabular}{cccccccc}
            \hline
            Name & $L_{\rm R}$ & $\Upsilon_{\rm R,h}$ & $\Upsilon_{\rm R,dm+h(<5R_{\rm eff})}$ & $\Upsilon_{\rm R,h}$ & $\Upsilon_{\rm R,h}$ & $\Upsilon_{\rm R,h}$ & $\Upsilon_{\rm R,Sal}$ \\
                 &  & Iso & Iso & New & Spl$_{a}$ & Spl & Sal \\
            (1) & (2) & (3) & (4) & (5) & (6) & (7) & (8)  \\
            \hline
            NGC\,821 & 1.87 & 3.74$\pm0.91$ & 23.4$^{+2.08}_{-1.98}$ & 12.5$\pm0.18$ & 10.5$\pm0.17$ & 10.9$\pm0.18$ & 5.92 \\
            NGC\,3379 & 1.62 & 1.11$\pm0.17$ & 11.7$^{+0.35}_{-0.34}$ & 5.49$\pm0.05$ & 5.19$\pm0.05$ & 5.31$\pm0.05$ & 6.88 \\
            NGC\,4374 & 5.87 & 2.25$\pm0.15$ & 15.6$^{+0.34}_{-0.34}$ & 7.87$\pm0.04$ & 7.34$\pm0.04$ & 7.44$\pm0.04$ & 7.01 \\
            NGC\,4697 & 2.31 & 6.49$\pm1.26$ & 7.36$^{+4.39}_{-0.72}$ & 6.75$\pm0.29$ & 6.06$\pm0.17$ & 6.19$\pm0.28$ & 6.31 \\
            \hline
            NGC\,1344 & 4.2  & 1.12$\pm0.05$ & 5.90$^{+0.09}_{-0.09}$ & 3.36$\pm0.12$ & 3.00$\pm0.12$ & 3.05$\pm0.12$ & $-$ \\
            NGC\,3379 & 1.62 & 1.11$\pm0.17$ & 11.7$^{+0.35}_{-0.34}$ & 5.43$\pm0.35$ & 5.00$\pm0.37$ & 5.19$\pm0.35$ & 6.88 \\
            NGC\,4374 & 5.87 & 2.25$\pm0.15$ & 15.6$^{+0.34}_{-0.34}$ & 7.94$\pm0.36$ & 7.26$\pm0.36$ & 7.44$\pm0.35$ & 7.01 \\
            NGC\,4494 & 2.48 & 2.82$\pm0.08$ & 6.66$^{+0.18}_{-0.18}$ & 4.40$\pm0.10$ & 3.59$\pm0.12$ & 3.75$\pm0.10$ & 5.0 \\
            NGC\,5846 & 4.62 & 6.06$\pm0.45$ & 17.8$^{+0.94}_{-0.91}$ & 10.7$\pm1.65$ & 9.29$\pm1.65$ & 9.59$\pm1.65$ & 7.13 \\
            \hline
            \end{tabular}
            \begin{minipage}{15.5cm}
            \textit{Notes.}
            (1) name of galaxy,
            (2) luminosity in R band with unit of $10^{10}L_{\odot}$ from \citep[ATLAS$^{\rm 3D}$,][]{Cappellari13a},
            (3) fitting R band mass-to-light ratio of of the galaxy in dark matter halo model
            (4) fitting R band mass-to-light ratio of dark matter halo and the galaxy within 5 effective radii,
            (5) fitting R band mass-to-light ratio of the galaxy in Newtonian gravity,
            (6) fitting R band mass-to-light ratio of the galaxy in MONDian gravity in simple form with anisotropic parameter $\beta$,
            (7) fitting R band mass-to-light ratio of the galaxy in MONDian gravity in simple form,
            (8) R band mass-to-light ratio estimated from population synthesis models with Salpeter IMF \citep[ATLAS$^{\rm 3D}$,][]{Cappellari13b}.
            All mass-to-light ratios are in unit of $M_{\odot}/L_{\odot}$.
            Columns (3) \& (4) are the results of two-parameter fitting in Newtonian gravity (luminous mass $M_{\rm h}$ and parameter of the isothermal dark matter halo $\sigma_v$).
            Columns (5), (6) \& (7) of first four galaxies (NGC\,4697, NGC\,821, NGC\,3379, NGC\,4374) are the results of one-parameter fitting
            (galaxy mass $M_{\rm h}$).
            Columns (5), (6) \& (7) of last five galaxies (NGC\,3379, NGC\,4374, NGC\,5846, NGC\,1344, NGC\,4494) are the results of two-parameter fitting
            (galaxy mass $M_{\rm h}$ and SMBH mass $M_{\rm bh}$).
            \end{minipage}
        \end{table*}

        In dark matter halo scenario, one expects that close to the center of a galaxy ordinary matter dominates dark matter,
        while at large distances from the center dark matter becomes dominant.
        Thus measurement of the dynamics at the outskirt of galaxies is crucial to dark matter scenario.
        As an alternative to dark matter scenario, MOND predicts the dynamics at the outskirt solely by the ordinary matter of the galaxy.
        We use PN.S data to trace the dynamics of 7 elliptical galaxies (see Table~\ref{tab:data}) up to 6--8 effective radii.
        Besides, we also use stellar data from SAURON~\citep{Coccato09} and ATLAS$^{\rm 3D}$~\citep{Cappellari11} to constraint the model at smaller radii.

        We adopt a spherically symmetric Hernquist model for these galaxies and allow for a SMBH at the center.
        We choose two interpolation functions in MOND in our study:
        standard form and simple form ($(\alpha,\eta)=(2,1)$ and $(1,1)$ in Equation~(\ref{eq:canonicalnu})).
        Moveover, we choose the acceleration parameter in MOND from the baryonic Tully-Fisher relation \citep{McGaugh11},
        i.e., $\mathfrak{a}_0=1.21\times 10^{-10}$ m s$^{-2}$.
        The characteristic length scale $r_{\rm h}$ is presumably provided by observed $R_{\rm eff}$ via $r_h\approx0.551\,R_{\rm eff}$.
        We thus have a two-parameter model (one for the mass of the galaxy $M_{\rm h}$ and the other the mass of the SMBH $M_{\rm bh}$),
        and if the mass of the SMBH is given then we have a one-parameter model.
        For comparison we also study dark matter scenario, in which we include an additional isothermal dark matter halo in Newtonian gravity
        (this is a two-parameter model, the mass of the galaxy $M_{\rm h}$ and $\sigma_v$ of the halo).

        The mass of the SMBH of 4 elliptical galaxies (NGC\,4697, NGC\,821, NGC\,3379 and NGC\,4374)
        in our sample has been measured by other method~\citep{McConnell11}.
        With a prescribed mass for the SMBH, we have only one parameter in our model for MOND, the mass of the galaxy $M_{\rm h}$.
        The result of the fitting by reduced $\chi$-square is shown in Figure~\ref{fig:fix}.

        Four models are shown in the figure:
        \begin{itemize}
        \item[(1)] dotted-dash line: Hernquist model with SMBH in MOND in simple form;
        \item[(2)] solid line: Hernquist model with SMBH in MOND in simple form with anisotropic parameter $\beta$ in Equation~(\ref{eq:anisotropic}),
        \item[(3)] dotted line: Hernquist model with SMBH in Newtonian gravity;
        \item[(4)] dash line: Hernquist model with isothermal dark matter halo (but without SMBH) in Newtonian gravity.
        \end{itemize}
        The first three are one-parameter models and the last one is a two-parameter model.
        The numerical values of the fitting parameter and the corresponding errors by reduced $\chi$-square method are shown in Table~\ref{tab:parameterfix} (also Table~\ref{tab:masstolight}).
        The table also lists the mass estimated by population synthesis.
        From the table, we learn that, except NGC\,821, the mass of other three elliptical galaxies matches well
        with the population synthesis model based on Salpeter IMF \citep{Cappellari13b}.
        We point out that the statistics of NGC\,821 is not very good due to insufficient stellar and PN data.

        When we let the mass of the SMBH as a free parameter, our model for MOND becomes a two-parameter model.
        There are 5 galaxies in our sample suitable for the analysis (NGC\,3379, NGC\,4374, NGC\,5846, NGC\,1344 and NGC\,4494).
        NGC\,4697 and NGC\,821 are not included because of the lack of stellar data in the inner region close to the center.
        We fit the data by reduced $\chi$-square and Figure~\ref{fig:two} shows the result of the 5 galaxies.
        Four models are shown in the figure.
        The four models are the same as in Figure~\ref{fig:fix} except that all four models are two-parameter models.
        The numerical values of the fitting parameters and the corresponding errors by reduced $\chi$-square method are listed in Table~\ref{tab:parametertwo} and Table~\ref{tab:masstolight}.
        From the table, we notice that the SMBH mass of NGC\,3379 and NGC\,4374 are consistent with the observed values \citep{McConnell11}.

        \begin{table*}
            \centering
            \setlength{\extrarowheight}{3pt}
            \caption[\textbf{Comparison with earlier works on NGC\,3379, NGC\,821 and NGC\,4494.}]
            {\textbf{Comparison with earlier works on NGC\,3379, NGC\,821 and NGC\,4494.}}\label{tab:comparison}
            \begin{tabular}{cc|ccccc|ccccc}
            \hline
                 &             & &\multicolumn{3}{c}{earlier works} & & & \multicolumn{3}{c}{present works} & \\
            Name & $L_{\rm B}$ & $D$ & $R_{\rm eff}$ & $N_{\rm PNe}$ & $\Upsilon_{\rm B,h}$ & $\xi$ & $D$ & $R_{\rm eff}$ & $N_{\rm PNe}$ & $\Upsilon_{\rm B,h}$ & $\xi$ \\
                 &             & Mpc & arcsec & & std & & Mpc & arcsec & & std & \\
            (1) & (2) & (3) & (4) & (5) & (6) & (7) & (8) & (9) & (10) & (11) & (12) \\
            \hline
            NGC\,821 & 2.47 &24 & 50 & 104 & 11.4 & 3.6 & 23.4 & 39.8 & 127 & 9.11$\pm0.14$ & 2.7 \\
            NGC\,3379 & 1.56 & 11 & 35 & 109 & 4.7 & 5.7 & 10.3 & 39.8 & 214 & 5.71$\pm0.05$ & 5.1 \\
            NGC\,4494 & 2.70 &17 & 49 & 73 & 5.4 & 3.4 & 16.6 & 50 & 267 & 3.93$\pm0.10$ & 3.0 \\
            \hline
            \end{tabular}
            \begin{minipage}{15.5cm}
            \textit{Notes.}
            (1) name of galaxy,
            (2) luminosity in B band with unit of $10^{10}L_{\odot}$ from \citet{Romanowsky03},
            (3) distance from \citet{Romanowsky03},
            (4) effective radius from \citet{Romanowsky03},
            (5) total number of PNe with measured radial velocities from \citet{Romanowsky03},
            (6) fitting B band mass-to-light ratio of the galaxy in MONDian gravity in standard form \citep{Milgrom&Sanders03},
            (7) the parameter $\xi$ of Equation~(\ref{eq:xi}) from \citet{Milgrom&Sanders03},
            (8) distance from ATLAS$^{\rm 3D}$ \citep{Cappellari11},
            (9) effective radius from ATLAS$^{\rm 3D}$ database \citep{Cappellari11},
            (10) total number of PNe with measured radial velocities from \citet{Coccato09},
            (11) fitting B band mass-to-light ratio of the galaxy in MONDian gravity in standard form,
            (12) the parameter $\xi$ of Equation~(\ref{eq:xi}) with mass estimated from population synthesis models with Salpeter IMF \citep{Cappellari13b}.
            All mass-to-light ratios are in unit of $M_{\odot}/L_{\odot}$.
            Columns (3) to (7) are the results of earlier works \citep[][]{Romanowsky03,Milgrom&Sanders03}.
            Columns (8) to (12) are the results of present work.
            \end{minipage}
        \end{table*}

        Our sample includes the three galaxies, NGC\,821, NGC\,3379 and NGC\,4494, studied by \citet{Romanowsky03}.
        In the past decade observation has improved a lot.
        There are about twice as many PNe data~\citep{Coccato09} and better measurement in distance and effective radius are obtained by
        ATLAS$^{\rm 3D}$~\citep{Cappellari11}.
        Our results and those of \citet{Romanowsky03} and \citet{Milgrom&Sanders03} on the three galaxies are consistent with each other.
        Table~\ref{tab:comparison} gives a detail comparison.
        Moreover, analysis on four other galaxies (NGC\,4374, NGC\,5846, NGC\,1344 and NGC\,3377) renders similar conclusion as in \citet{Romanowsky03},
        Newtonian dynamics is close to data of velocity dispersion within 6 effective radius.
        This runs into the tenet of dark matter halo models which are supposed to have more dark matter at the outer part of galaxies.

        Hernquist model in MOND fits the data well (see Figures~\ref{fig:fix} \& \ref{fig:two} and Table~\ref{tab:masstolight}).
        The existence of a SMBH has almost no influence on the dynamics at distances beyond one effective radius
        (the same conclusion as the case for pure Newtonian gravity discussed in Section\ref{sec:Hernquist}, see Figure~\ref{fig:BH}).
        In our samples, different interpolation functions (such as simple form and standard form) only produce very small difference
        in the profile of the velocity dispersion and give very similar best fit parameters.

        All the masses of the galaxies found under MOND are at most 17\% less than those found under Newtonian gravity.
        The mass discrepancy is small because the ``effective acceleration'' is large comparing with $\mathfrak{a}_0$, the acceleration parameter of MOND,
        thus the systems are closer to Newtonian regime.
        Recall the parameter $\xi$ from Equation~(\ref{eq:xi})
        $\xi=\sqrt{GM/R_{\rm eff}^2\mathfrak{a}_0\,}$.
        We calculate $\xi$ by taking the mass estimated by population synthesis with Salpeter IMF and
        the effective radius in ATLAS$^{\rm 3D}$ \citep{Cappellari11},
        and choose $\mathfrak{a}_0=1.21\times 10^{-10}$ m s$^{-2}$ \citep{McGaugh11}.
        We list $\xi$ and the mass (from Newtonian gravity and MOND) in Tables~\ref{tab:parameterfix} and \ref{tab:parametertwo}.
        All of them are larger than unity ($\xi\sim 2.7$ to $5.3$), cf. HSB spiral galaxies (see the end of Section~\ref{sec:MOND}).
        Large $\xi$ is indicative of system close to Newtonian regime, and therefore, the difference in masses found by Newtonian theory
        and MOND is expected to be small.

        To compare with dark matter model, we consider a Hernquist model for the galaxy and a singular isothermal sphere for the dark matter halo
        in Newtonian gravity.
        The dark matter model can fit the data as well.
        However, for some cases such as NGC\,3379, NGC\,4374 and NGC\,4494,
        the baryonic mass of the galaxy is a lot less than expected from population synthesis models with Salpeter IMF,
        see Tables~\ref{tab:parameterfix} and \ref{tab:parametertwo}.
        Within the dark matter halo model five galaxies (NGC\,821, NGC\,3379, NGC\,4374, NGC\,5846 and NGC\,1344) have dark halo mass $M_{\rm dm}$
        (mass of dark matter within $5R_{\rm eff}$) significantly larger than the baryonic mass $M_{\rm h}$.
        However, these $M_{\rm dm}$ are not much larger than the Salpeter mass.
        We would like to point out that NGC\,4697 has negligible dark matter and the best fit is very close to the model without dark halo (see Figure~\ref{fig:fix}).
        In Table~\ref{tab:masstolight}, one notices that the
        mass-to-light ratios of NGC\,3379, NGC\,4374, NGC\,1344 and NGC\,4494 are too small for common dark matter model.

        \begin{figure}
            \centering
            \includegraphics[width=\columnwidth]{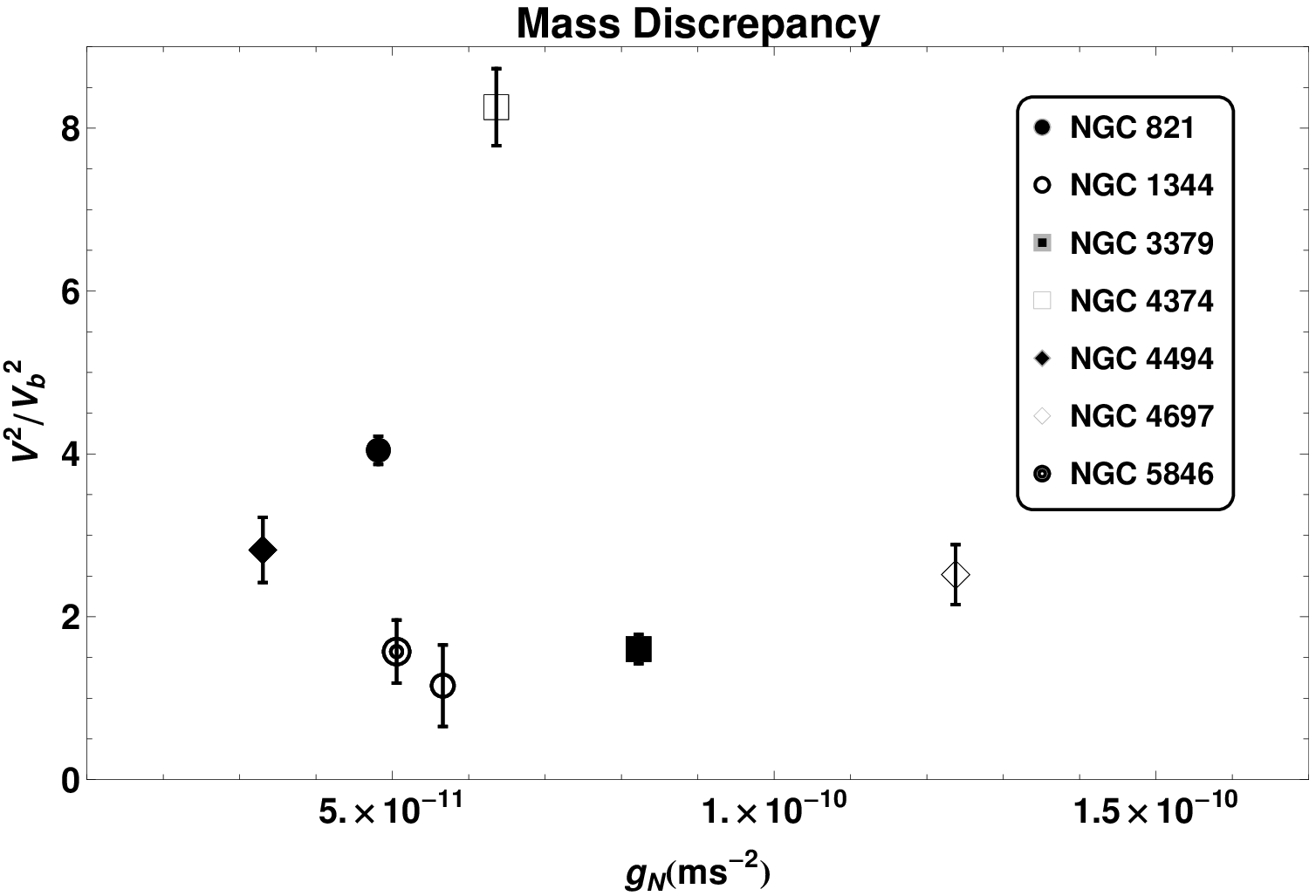}
            \caption{Mass discrepancy (see text for definition) is larger for smaller gravitational acceleration.
            Horizontal axis is the Newtonian acceleration at the location of the furthest PN by the stellar mass of the galaxy
            \citep[Salpeter initial mass function is used,][]{Cappellari13b}.
            Error in mass discrepancy comes from error in the velocity dispersion of PNe.
            }
            \label{fig:MassDis}
        \end{figure}

        In addition, our work confirms two predictions by MOND: (1) mass discrepancy increases as Newtonian acceleration decreases
        \citep{McGaugh04,FM12}, and (2) maximum halo acceleration \citep{Milgrom&Sanders05,Gentile09}.

        Mass discrepancy refers to the ratio of the dynamical mass to the baryonic mass.
        Usually it is represented by $V^2/V^2_{\rm b}$, where $V$ is the circular velocities by the total mass of the galaxy and $V_{\rm b}$ is
        the circular velocity by the baryonic mass.
        From the measured velocity dispersion $\sigma_I$ at a distance $R$ far away from the centre of galaxy, we can estimate $V$ by putting $g=V^2/r$
        and the corresponding $\rho=V^2/4\pi Gr^2$ into Equations~(\ref{eq:Jeans}) \& (\ref{eq:vel_I}) (with $\beta=0$) \citep[][]{Churazov10}.
        $V_{\rm b}$ is estimated by $V^2_{\rm b}=GM_{\rm b}/R$ \citep[][]{McGaugh04}, and $M_{\rm b}$ is obtained by population synthesis
        \citep[using Salpeter initial mass function][]{Cappellari13b}.
        Figure~\ref{fig:MassDis} shows the mass discrepancy (represented by $V^2/V^2_{\rm b}$) of our sample on elliptical galaxies.
        This is similar to the result on spiral galaxies \citep[e.g.,][]{McGaugh04,FM12}.
        Recently, \citet{Janz16} used data from ATLAS$^{\rm 3D}$ and SLUGGS on the problem of mass discrepancy.
        Our result is consistent with theirs.

\begin{figure}
            \centering
            \includegraphics[width=\columnwidth]{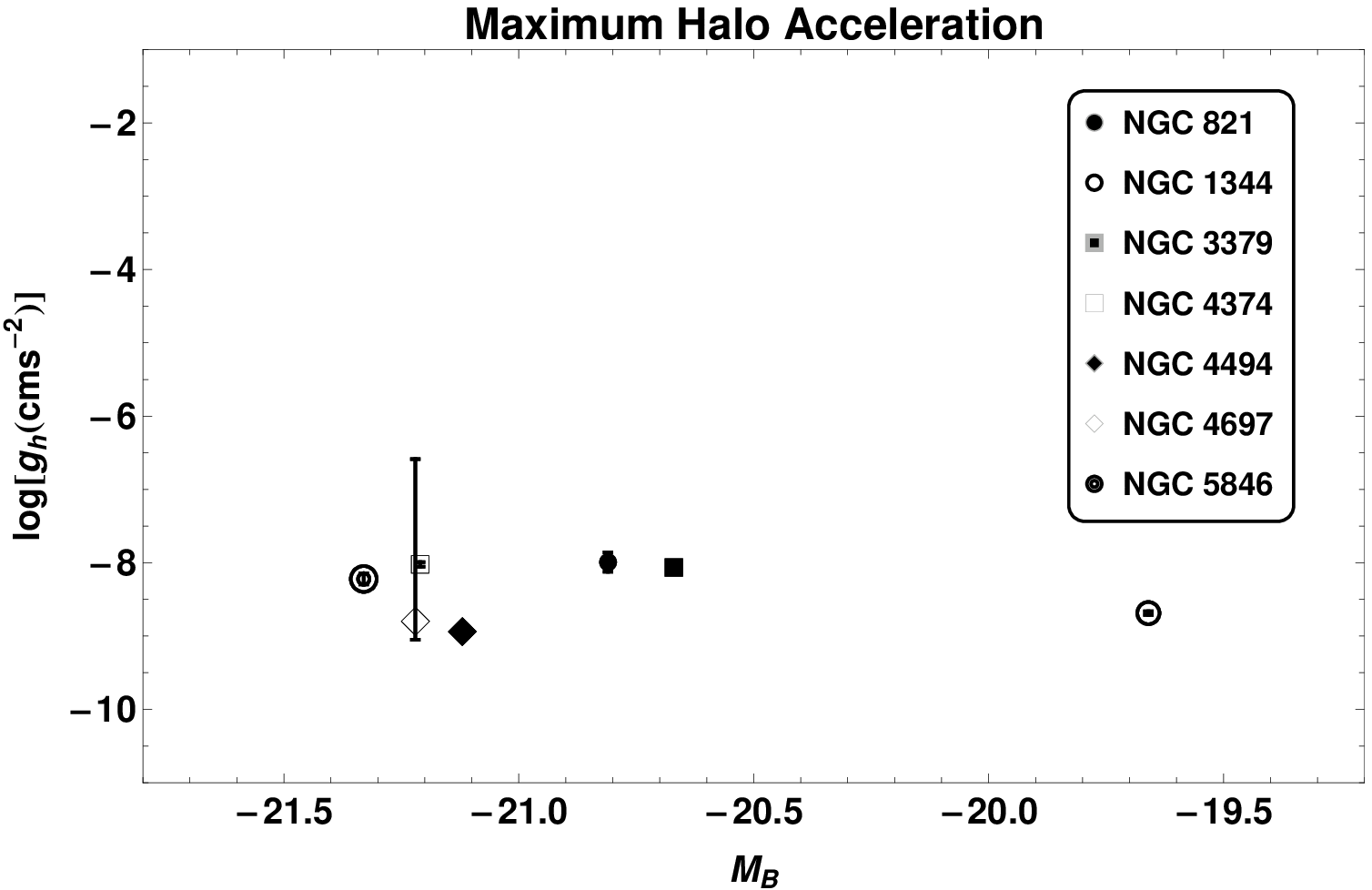}
            \includegraphics[width=\columnwidth]{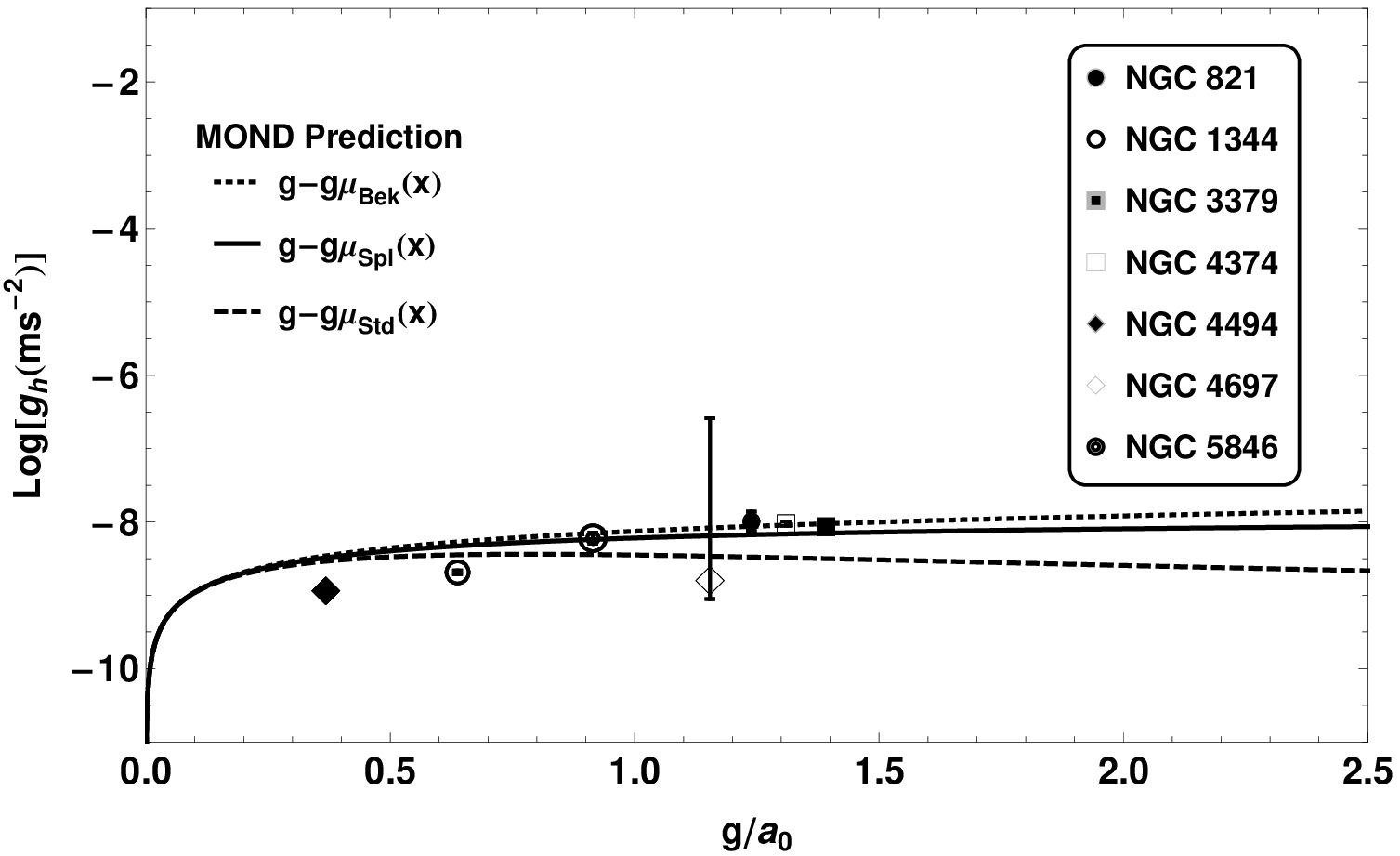}
            \caption{Upper panel: Maximum halo acceleration is the difference between the MONDian and Newtonian gravitational acceleration.
            The horizontal axis is the absolute B-band magnitude of the galaxies (see Table~\ref{tab:data}).
            The error in $g_{\rm h}$ comes from the error in the velocity dispersion of PNe.
            Except for NGC\,4697, the error bar in the figure is smaller than the symbol.
            Lower panel: The horizontal axis is $x=g/a_\mathfrak{0}=(g_{\rm N}+g_{\rm h})/a_\mathfrak{0}$.
            The dotted line, solid line and dashed line correspond to Bekenstein form, simple form and standard form, respectively.
            }
            \label{fig:mha}
        \end{figure}

        In dark matter scenario, when we use isothermal distribution for the dark matter halo and Hernquist model (with Salpeter mass) for the galaxy
        to fit the velocity dispersion,
        we get the acceleration contributed by the halo $g_{\rm h}$ and by the galaxy $g_{\rm N}$.
        The result is consistent with \citet{Gentile09}, see upper panel of Figure~\ref{fig:mha}.
        In MOND's interpretation, $g_{\rm h}=g-g_{\rm N}=g(1-\mu(g/a_\mathfrak{0}))$ \citep{Milgrom&Sanders05}.
        We compare the prediction of MOND with the result from dark matter halo in the lower panel of Figure~\ref{fig:mha}.

        In summary, MOND can naturally explain the dynamics of the 7 galaxies listed in Table~\ref{tab:data} up to 6 effective radii.
        Since the parameter $\xi$ is large in these galaxies
        (i.e., the characteristic acceleration is large compare to $\mathfrak{a}_0$),
        the acceleration discrepancy (or mass discrepancy) is small.
        MOND is close to Newtonian regime in these cases
        (or if dark matter halo model is used, little dark matter is needed up to several effective radii).
        Consequently, only slightly different results are obtained when different (inverted) interpolation functions $\tilde{\nu}(x)$ are used.
        We have also considered interpolation functions other than simple form.
        For instances, for the standard form, the best fitted mass of a galaxy is larger than that of simple form by $10^{+4}_{-6}\%$.
        Once again, this shows that $\mathfrak{a}_0$ is an important parameter (if not the most)
        to distinguish between Newtonian regime and MONDian regime (or dark matter dominated regime).
        It is imperative to search for high quality data for elliptical galaxies with small $\xi$,
        so that one can perform study similar to the study on HSB and LSB spiral galaxies \citep{Sanders&McGaugh02}.

The authors are grateful to N. Kanekar for stimulating discussion on this work
and to M. Milgrom for giving many helpful comments and advices.
We especially thank the anonymous reviewer for the comments and suggestions to improve our work.
We also thank D. H. Hartmann and C. H. Chang for giving useful information on statistical methods of regression.
This work is supported in part by the Taiwan Ministry of Science and Technology
grants MOST 102-2112-M-008-019-MY3 and MOST 104-2923-M-008-001-MY3.


\begin{thebibliography}{}
\bibitem[Bekenstein \& Milgrom(1984)]{BM84}
Bekenstein, J. D., Milgrom, M., 1984, ApJ, 286, 7
\bibitem[Bekenstein(2004)]{TeVeS}
Bekenstein, J. D., 2004, Phys.Rev. D70, 083509
\bibitem[Binney \& Tremaine(2008)]{Binney2008}
Binney, J., Tremaine, S. 2008, Galactic Dynamics (2nd ed.), Chapter 4, Princeton University Press
\bibitem[Cappellari et al.(2011)]{Cappellari11}
Cappellari, M., Emsellem, E., Krajnovic D., McDermid R. M., et al., 2011, MNRAS, 413, 813
\bibitem[Cappellari et al.(2013a)]{Cappellari13a}
Cappellari, M., McDermid, R. M., Alatalo, K., et al., 2013a, MNRAS, 432, 1709
\bibitem[Cappellari et al.(2013b)]{Cappellari13b}
Cappellari, M., McDermid, R. M., Alatalo, K., et al., 2013b, MNRAS, 432, 1862
\bibitem[Chiu et al.(2006)]{ckt06}
Chiu, M. C., Ko, C. M., Tian, Y., 2006, ApJ, 636, 565
\bibitem[Churazov et al.(2010)]{Churazov10}
Churazov, E., Tremaine, S., Forman, W., Gerhard, O., 2010, MNRAS, 404, 1165
\bibitem[Chiu et al.(2011)]{ckt11}
Chiu, M. C., Ko, C. M., Tian, Y., Zhao, H. S., 2011, Phys.Rev. D83, 063523
\bibitem[Coccato et al.(2009)]{Coccato09}
Coccato, L., et al. 2009, MNRAS, 394, 1249
\bibitem[de Lorenzi et al.(2007)]{de Lorenzi07}
de Lorenzi, F., Debattista, V. P., Gerhard, O., et al., 2007, MNRAS, 376, 71
\bibitem[de Lorenzi et al.(2009)]{de Lorenzi09}
de Lorenzi, F., Gerhard, O., Coccato, L., et al., 2009, A\&{A}, 502, 771
\bibitem[Douglas et al.(1997)]{Douglas97}
Douglas, N. G., Taylor, K., Freeman, K. G., et al., 1997, IAUS Symposium, Vol. 180, Planetary Nebulae, ed. Habing, H. J., Lamers, H. J., 493
\bibitem[Douglas et al.(2002)]{Douglas02}
Douglas, N. G., Arnaboldi, M., Freeman, K. G., et al. 2002, PASP~114, 801
\bibitem[Douglas et al.(2007)]{Douglas07}
Douglas, N. G., Napolitano, N. R., Romanosky, A. J., et al. 2007, ApJ, 664, 265
\bibitem[Famaey et al.(2007)]{Famaey07}
Famaey, B., Bruneon. J.-P., Zhao, H. S., 2007, MNRAS, 377, L79
\bibitem[Famaey \& Binney(2005)]{FB05}
Famaey, B., Binney, J., 2005, MNRAS, 363, 603
\bibitem[Famaey \& McGaugh(2012)]{FM12}
Famaey, B., McGaugh, S. S., 2012, Living Rev. Relativity, 15, 10
\bibitem[Gentile et al.(2009)]{Gentile09}
Gentile, G., Famaey, B., Zhao, H. S., Salucci, P., 2009, Nature, 461, 627
\bibitem[Hernquist(1990)]{Hernquist90}
Hernquist, L., 1990, ApJ, 356, 359
\bibitem[Hui et al.(1995)]{Hui95}
Hui, X., Ford, H. C., Freeman, K. C., et al., 1995, ApJ, 449, 592
\bibitem[Janz et al.(2016)]{Janz16}
Janz, J., Cappellari, M., Romanowsky, A. J., et al., 2016, arXiv:1606.05003
\bibitem[Lubin et al.(2000)]{Lubin00}
Lubin, L. M., et al., 2000, AJ, 119, 451
\bibitem[Makarov et al.(2009)]{Makarov09}
Makarov, D., Prugniel, P., Terekhova, N., et al., 2014, A\&{A}, 570, 13
\bibitem[McConnell et al.(2011)]{McConnell11}
McConnell, N. J., Ma, C. P., Gebhardt, K., et al. 2011, Nature, 480, 215
\bibitem[Milgrom(1983)]{Milgrom83}
Milgrom, M., 1983, ApJ, 270, 365
\bibitem[Milgrom(1983)]{Milgrom83b}
Milgrom, M., 1983, ApJ, 270, 371
\bibitem[Milgrom \& Sanders(2003)]{Milgrom&Sanders03}
Milgrom, M., Sanders, R. H., 2003, ApJL, 599, L25
\bibitem[Milgrom \& Sanders(2005)]{Milgrom&Sanders05}
Milgrom, M., Sanders, R. H., 2005, MNRAS, 357, 45
\bibitem[Milgrom(2009)]{Milgrom09}
Milgrom, M. 2009, Phys.Rev. D80, 123536
\bibitem[Milgrom(2012)]{Milgrom12}
Milgrom, M. 2012, Phys. Rev. Lett., 109, 131101
\bibitem[McGaugh(2004)]{McGaugh04}
McGaugh, S. S., 2004, ApJ, 609, 652
\bibitem[McGaugh(2011)]{McGaugh11}
McGaugh, S. S., 2011, Phys. Rev. Lett., 106, 121303
\bibitem[Napolitano et al.(2009)]{Napolitano09}
Napolitano, N. R., Romanowsky, A. J., Coccato, L., et al. 2009, MNRAS, 395, 76
\bibitem[Richtler et al.(2008)]{Richtler08}
Richtler, T., Schuberth Y., Hilker M., et al., 2008, A\&{A}, 478, L23
\bibitem[Riess et al.(2011)]{Riess11}
Riess, A. G., et al., 2011, ApJ, 730, 113
\bibitem[Rodrigues(2012)]{Rodrigues12}
Rodrigues, D. C., 2012, J. Cosmol. Astropart. Phys., 09, 031
\bibitem[Romanowsky et al.(2003)]{Romanowsky03}
Romanosky, A. J., Douglas, N. G., Arnaboldi, M., et al., 2003, Science, 301, 1696
\bibitem[Rubin \& Fort(1970)]{Rubin}
Rubin, V., Fort, W. K. 1970, ApJ, 159, 379
\bibitem[Samurovi\'{c}(2012)]{Samurovic12}
Samurovi\'{c}, S., 2012, A\&{A}, 541, A138
\bibitem[Samurovi\'{c}(2014)]{Samurovic14}
Samurovi\'{c}, S., 2014, A\&{A}, 570, A132
\bibitem[Samurovi\'{c} \& \'{C}irkovi\'{c}(2008)]{Samurovic&Cirkovic08}
Samurovi\'{c}, S., \& \'{C}irkovi\'{c}, M. M., 2008, A\&{A}, 488, 873
\bibitem[Sanders \& McGaugh(2002)]{Sanders&McGaugh02}
Sanders, R. H., McGaugh, S. S., 2002, ARA{\&}A, 40, 263
\bibitem[Sanders(2014)]{Sanders14}
Sanders, R. H., 2014, MNRAS,439, 1781
\bibitem[Schuberth et al.(2012)]{Schuberth12}
Schuberth Y., Richtler, T., Hilker M., et al. 2012, A\&{A}, 544, A115
\bibitem[Tian et al.(2013)]{Tian13}
Tian, Y., Ko, C. M., Chiu, M. C. 2013, ApJ, 770, 154
\bibitem[Teodorescu et al.(2005)]{Teodorescu05}
Teodorescu, A. M., Mendez, R. H., Saglia, R. P., et al. 2005, ApJ, 635, 290
\bibitem[Tian et al.(2013)]{Tian13}
Tian, Y., Ko, C. M., Chiu, M. C., 2013, ApJ, 770, 154
\bibitem[Tortora et al.(2014)]{Tortora14}
Tortora, C., Romanowsky, A. J., Cardone, V. F. et al., 2014, MNRASL, 438, L46
\bibitem[van Albada(1982)]{vanAlbada82}
van Albada, T. S., 1982, MNRAS, 201, 939
\bibitem[Zhao(2006)]{Zhao06}
Zhao, H. S., Bacon, D. J., Taylor, A. N., Horne, K., 2006, MNRAS, 368, 171
\bibitem[Zlosnik et al.(2007)]{Zlosnik07}
Zlosnik, T. G., Ferreira, P. G., Starkman, G. D. 2007, Phys.Rev. D75, 044017
\bibitem[Zwicky(1933)]{Zwicky}
Zwicky, F., 1933, Helv. Phys. Acta~6, 110
\end{thebibliography}
\end{document}